\documentclass[secnumarabic,superscriptaddress,amssymb, aps,  prb,twocolumn]{revtex4-2}

\usepackage{amsmath}
\usepackage{mathrsfs}  
\usepackage{mathtools} 
\usepackage{bbold}   
\usepackage{braket}    
\usepackage{graphicx}
\usepackage{threeparttable}
\usepackage{amssymb}
\usepackage{gensymb} 
\usepackage{graphicx}     
\usepackage{wrapfig}
\usepackage{multirow}    
\usepackage{here} 		
\usepackage{url}
\usepackage{soul}               
\usepackage{xspace}
\usepackage{booktabs}  
\usepackage{titlesec}
\usepackage{newunicodechar}
\usepackage{placeins}
\usepackage{float}  
\usepackage{bibunits}
\usepackage{siunitx}
\usepackage{setspace}
\usepackage{tabularx}
\usepackage{array}
\usepackage{rotating}
\usepackage{makecell}
\usepackage{adjustbox}
\usepackage[table,dvipsnames]{xcolor}
\usepackage[
    plainpages=false,
    pdfpagelabels=true,
    linkcolor=cyan,
    breaklinks,
    colorlinks=true
]{hyperref}

\DeclareUnicodeCharacter{03B1}{\ensuremath{\alpha}}

\makeatletter

\let\auto@bib@innerbib\@empty

\AddToHook{begindocument/end}{%

  \IfHookExistsTF{cmd/hyper@natlinkstart/before}{%
    \RemoveFromHook{cmd/hyper@natlinkstart/before}[hyperref]%
  }{}%

  \IfHookExistsTF{cmd/hyper@natlinkend/after}{%
    \RemoveFromHook{cmd/hyper@natlinkend/after}[hyperref]%
  }{}%

  \def\hyper@natlinkstart#1{%
    \Hy@backout{#1}%
    \hyper@linkstart{cite}{cite.bu\the\@bibunitauxcnt.#1}%
    \def\hyper@nat@current{#1}%
  }%

  \def\hyper@natlinkend{\hyper@linkend}%

  \def\hyper@natlinkbreak#1#2{%
    \hyper@linkend#1%
    \hyper@linkstart{cite}{cite.bu\the\@bibunitauxcnt.#2}%
  }%

  \def\hyper@natanchorstart#1{%
    \Hy@raisedlink{%
      \hyper@anchorstart{cite.bu\the\@bibunitauxcnt.#1}%
    }%
  }%

  \def\hyper@natanchorend{\hyper@anchorend}%
}

\makeatother

\newcommand{\kim}[1]{{\color{Purple}  #1}}
\newcommand{\ar}{$\alpha$-RuCl$_3$\xspace}
\newcommand{\rucl}{RuCl$_3$\xspace}

\newcommand{\cstar}{$c^\star$\xspace}

\newcommand{\mono}{\textit{C2/m}\xspace}
\newcommand{\rhom}{\textit{$R\bar{3}$}\xspace}

\begin{document}
\begin{bibunit}[apsrev4-2]

\title{A thermodynamic reference for field-induced phenomena in \ar: two coexisting polymorphs account for multiple magnetic phase transitions}

\author{Hamza Nasir}
\affiliation{Institute of Science and Technology Austria, 3400 Klosterneuburg, Austria}
\author{Daniel Balazs}
\affiliation{Institute of Science and Technology Austria, 3400 Klosterneuburg, Austria}
\author{Muhammad Nauman}
\affiliation{Institute of Science and Technology Austria, 3400 Klosterneuburg, Austria}
\author{Vikram Nagarajan}
\affiliation{Institute of Science and Technology Austria, 3400 Klosterneuburg, Austria}
\author{Ezekiel Horsley}
\affiliation{Department of Physics, University of Toronto, Toronto, Ontario, Canada M5S 1A7}
\author{Subin Kim}
\affiliation{Department of Physics, University of Toronto, Toronto, Ontario, Canada M5S 1A7}
\author{Young-June Kim}
\affiliation{Department of Physics, University of Toronto, Toronto, Ontario, Canada M5S 1A7}
\author{K.~A.~Modic}
\email{kimberly.modic@ist.ac.at}
\affiliation{Institute of Science and Technology Austria, 3400 Klosterneuburg, Austria}

\date{\today}
\begin{abstract}
The majority of research on $\alpha$-RuCl$_3$ has focused on applying in-plane magnetic fields to suppress the antiferromagnetic order and induce a quantum spin liquid (QSL). However, interpreting field-induced phenomena has been complicated by the material's temperature- and sample-size-dependent crystal structure and its sensitivity to strain-induced stacking disorder. Here, we isolate the monoclinic ($C2/m$) phase in nanogram-scale single crystals, providing the opportunity to study Kitaev physics in a new setting. We perform high-resolution magnetotropic susceptibility measurements on a structurally well-defined crystal within several crystal planes. Mapping the antiferromagnetic (AFM) phase boundary as a function of temperature, magnetic field, and field orientation, we find that the monoclinic phase diagram closely resembles that of rhombohedral crystals, but is systematically shifted to higher transition temperatures and critical fields. We find that the critical fields coincide quantitatively with anomalies observed across the $\sim$7-11~T field range, where Kitaev spin liquid physics has been widely reported across numerous independent studies and experimental techniques. Our study shows that these anomalies arise from the magnetic phase boundaries of coexisting monoclinic and rhombohedral structural domains, rather than an exotic phase. These results provide a thermodynamic reference against which past studies can determine their structural composition, and toward which future studies can look for genuine signatures of spin-liquid phenomena beyond the close vicinity of the AFM boundary under in-plane fields.
\end{abstract}



\maketitle

\section{Introduction}

The layered honeycomb magnet \ar (hereafter \rucl) is a leading candidate for realizing a Kitaev quantum spin liquid (QSL) \cite{Kitaev2006, Jackeli2009, Kim2022, Loidl2021, ojeda2025lessons}. In magnetic fields applied within the honeycomb plane, the zero-field antiferromagnetic (AFM) order is suppressed between $\sim$7-11 T, depending upon sample structure and quality. Above this critical field, reports of a quantized thermal Hall conductance \cite{Kasahara2018a} and oscillatory thermal transport \cite{Czajka2021, Bruin2023, suetsugu2022evidence, Lefrancois2023, Xing2025, Bruin2022} have been interpreted as possible fingerprints of fractionalized, charge-neutral excitations. The origin of these transport anomalies has been debated for several years: some studies attribute them to an intrinsic spin-liquid state, while others attribute them to disorder in the material's stacking of honeycomb layers---but no study has pinned down the underlying mechanism. 

The difficulty in resolving this question stems from \rucl's sensitivity to stacking order and a strain-dependent structural transformation. Near $T \approx 150$ K, the material transforms from its high-temperature monoclinic ($C2/m$) structure \cite{johnson2015monoclinic, Cao2016} into a rhombohedral ($R\bar{3}$) structure upon cooling \cite{Kubota2015}. However, this structural transformation depends sensitively on sample size and quality \cite{Kim2024}: larger crystals with lower N\'{e}el temperatures ($T_\text{N} \sim 7$ K) typically transform into the \rhom structure at low temperature, whereas smaller crystals with higher transition temperatures ($T_\text{N} \sim 14$ K) remain \mono \cite{Kim2024}.

Most previous studies have focused on larger crystals that are predominantly \rhom, and which have historically been regarded as higher quality. However, recent measurements---including magnetic anisotropy, thermal transport, and diffuse x-ray scattering---suggest that \mono domains may persist even in samples believed to be predominantly rhombohedral and of high quality at room temperature \cite{Sears2023, Zhang2024, Kim2024}. 

Upon cooling, surface strain effects may further stabilize \mono regions, leading to samples with coexisting \rhom and \mono structural domains at low temperatures \cite{Kim2025}. This possibility, as well as an extreme sensitivity to stacking disorder \cite{Cao2016}, raises the prospect that some reported field-induced phenomena may have a structural origin.

Indeed, \citet{Lefrancois2023} suggested that the thermal transport anomalies likely arise from magnetic instabilities associated with stacking disorder. More recently, studies using crystals with reduced stacking disorder \cite{Xing2025, Zhang2024} found that the anomalies are attenuated in cleaner samples. However, the low-temperature structure is difficult to determine directly, and thermal transport measurements offer no means of isolating or verifying a single structural domain in the region actually probed---a difficulty common to 2D layered samples where strain and sample handling tend to introduce additional domains. The precise origin of the anomalies has therefore remained undetermined even in these more carefully controlled studies.

Here, we isolate a single structural domain of \mono symmetry at low temperature---the polymorph rarely reported on---and map its AFM phase boundary across temperature, magnetic field, and field orientation using high-resolution magnetotropic susceptibility measurements \cite{Modic2018, Shekhter_2023}.
We find that the \mono phase diagram closely resembles that of the \rhom phase, including a field-induced intermediate ordered phase analogous to the $ZZ2$ phase found in the \rhom structure \cite{Sears2017, Wagner2022}, which is confined to within $\sim30^\circ$ of the $a$-axis in both structures. 



We show that the two critical fields of the AFM phases for $B||a$ associated with the \mono structure trace to the two higher-field oscillatory features observed in thermal transport experiments \cite{Czajka2021, Bruin2023, suetsugu2022evidence, Lefrancois2023}. Combining this finding with the well-known AFM transitions found in nominally \rhom samples \cite{Sears2017, Balz2019, Balz2021, Tanaka2022, Wagner2022, bachus2020thermodynamic}, we show that the magnetic phase transitions of both coexisting polymorphs---naturally present in most samples---coincide with the field values of the reported transport anomalies, providing a magnetic, structurally-specific origin distinct from the generic stacking disorder invoked in prior work \cite{Lefrancois2023,Bruin2023,Xing2025,Zhang2024}. This structurally-grounded framework extends beyond the oscillatory transport anomalies to nearly the full body of field-induced transitions reported in \rucl, as we establish by comparing transitions identified in twelve independent studies in Table~\ref{tab:tn_bc_compilation}.

\begin{figure}[ht!]
  \centering
  \includegraphics[width=0.7\columnwidth]{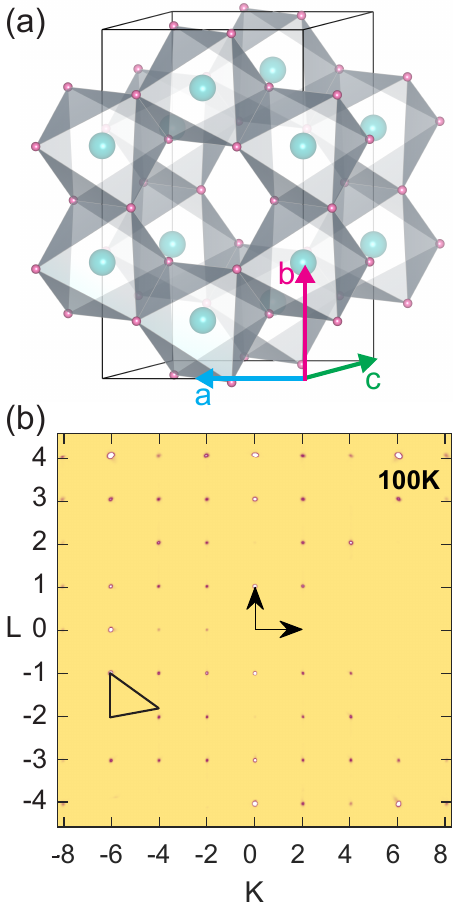}
  \caption{(a) Crystal structure of monoclinic (\mono space group) \rucl, highlighting the layered honeycomb arrangement of Ru atoms formed by edge-shared Cl octahedra.
(b) Reciprocal-space map in the ($0kl$) plane obtained from single-crystal X-ray diffraction on Sample S2 at 100 K. Bragg reflections occur at integer reciprocal-lattice positions, indicating that the crystal remains in the monoclinic (\mono) phase down to 100 K. The triangle highlights the absence of intensity at reciprocal-space positions corresponding to the rhombohedral phase (at $l = n + k/6$, where $n$ is an integer).
}
  \label{fig:figure1}
\end{figure}

\section{Experiment}
High-quality \rucl single crystals were grown by the chemical vapor transport method described in Ref. \cite{Kim2022}. Four single crystals from the same growth batch were used for our measurements, each selected according to the experimental requirements: Sample S1, having the dimensions of $70 \times 55 \times 3~\mu$m (approximately 35 ng) was used for the magnetotropic susceptibility measurements. Sample S2 of comparable size was selected for single-crystal X-ray diffraction (SC-XRD). Other samples measured to confirm reproducibility are presented in the SI. All sample information can be found in the SI \S IV. 
%


For magnetotropic measurements, Sample S1 was carefully transferred onto a 3~$\mu$m thick piece of silicon. This sample size is a deliberate choice rather than a limitation: resonant torsion magnetometry (RTM) is sensitive enough to resolve magnetic phase transitions on crystals far below the mass required for heat capacity, thermal conductivity, neutron diffraction, or ultrasound measurements, and it is precisely at this scale that a single structural domain of $C2/m$ symmetry can be isolated. 

\begin{figure*}[ht]
  \centering
  \includegraphics[width=0.8\textwidth, ]{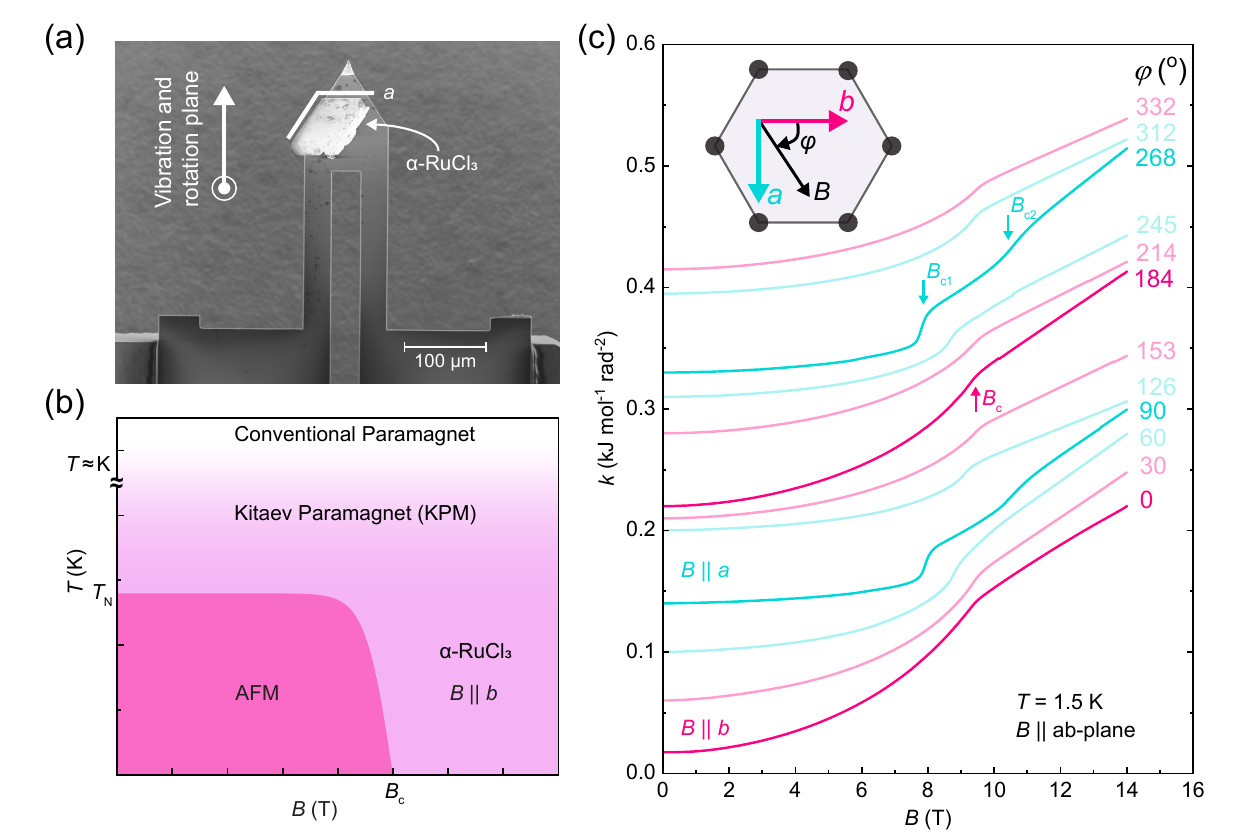}
  \caption{
  (a) Scanning electron microscopy image of a small \rucl crystal mounted on the tip of a silicon cantilever. The in-plane crystallographic $a$-axis is shown. The vibration and rotation plane of the cantilever is shown by the arrow to the left of the cantilever.
  (b) A schematic temperature-field ($T$–$B$) phase diagram showing the antiferromagnetic phase boundary of \rucl (dark pink region). At higher temperatures and fields, the system enters the Kitaev-paramagnetic (KPM) regime (light pink), followed by a conventional paramagnetic state at elevated temperatures (white region).
  (c) Magnetotropic susceptibility as a function of magnetic field applied within the honeycomb plane. The inset (upper left) defines the in-plane field angle relative to the crystallographic axes.
  }
  \label{fig:figure2}
  \end{figure*}

The crystal was attached to the silicon with a small drop of Apiezon N grease, which hardens under exposure to the electron beam in the SEM. After securing the sample to silicon, we used focused-ion beam (FIB) milling to cut through the silicon around the crystal (SI Fig.~S4 and \autoref{fig:figure2}(a)). The crystal and the silicon below it were transferred to a silicon cantilever, known as the Akiyama A-probe \cite{Akiyama2010}, for resonant torsion magnetometry (RTM) experiments \cite{Modic2018, Shekhter_2023} (SI \S III).

In between each measurement ($\textit{i.e.}$ after complete rotation of the magnetic field in a single crystal plane that includes the $c^\star$-axis), an eyelash was used to push the side of the silicon piece under the crystal in order to rotate it. This method minimizes the potential to damage the sample.

RTM detects the magnetotropic susceptibility $k$---the second-derivative of the free energy $F$ with respect to magnetic field orientation $\theta$, $k = \partial^2F/\partial \theta ^2$ \cite{Modic2018, Shekhter_2023}. Data on \rucl were collected while systematically rotating the magnetic field in 12 crystallographic planes at various fixed magnetic fields and temperatures. $\theta$ is defined as the angle between the magnetic field and the $c^\star$-axis (perpendicular to the honeycomb planes), and $\phi$ describes the azimuthal angle---rotation of the crystal around the $c^\star$-axis with $\phi = 0$ corresponding to $B||b$. $\phi$ was changed in $\sim$30$^\circ$ increments spanning from $\phi =0^\circ$ to 332$^\circ$. Field sweeps up to 14~T were performed at base temperature (1.5~K), while temperature sweeps between 20~K and 1.5~K were carried out at fixed fields ranging from 2 to 14~T.

\section[res]{Results}
\autoref{fig:figure1}(a) shows the \mono crystal structure of \rucl, highlighting the layered honeycomb arrangement of Ru atoms coordinated by edge-shared Cl octahedra \cite{Kubota2015}. The monoclinic unit cell is outlined, emphasizing that the honeycomb planes are stacked perpendicular to the $ab$-plane, and shifted along the $a$-direction.

To determine the structural symmetry of Sample S2, we created 2D reciprocal space maps (RSMs) that highlight the stacking order at both room temperature and low temperature (SI Fig.~S1). \autoref{fig:figure1}(b) shows an RSM in the 
$(0kl)$ plane at 100~K. Following the procedure reported by \citet{Kim2024}, the Miller indices $h$, $k$, and $l$ refer to a rectilinear basis, where the $a$- and $b$-axes are preserved and the $c$-axis is defined along the reciprocal lattice   $c^\star$. Since the neighboring layers in the \mono phase are shifted along $a$, a perfect rectangular grid is observed in this representation. Peaks associated with phases where the shift along $b$ is non-zero, such as the rhombohedral phase, will appear at non-integer positions indicated by the triangle in \autoref{fig:figure1}(b). However, at both 300~K and 100~K, the RSMs in the rectilinear $(0kl)$ plane dominantly display Bragg reflections forming a rectangular grid at integer $(hkl)$  coordinates, consistent with monoclinic stacking \cite{Kubota2015} (SI \S I.1). From our structural analysis, we estimate that the fraction of \rhom volume is of the order less than 1\%.

\begin{figure*}[ht]
  \centering
  \includegraphics[width=0.8\textwidth,]{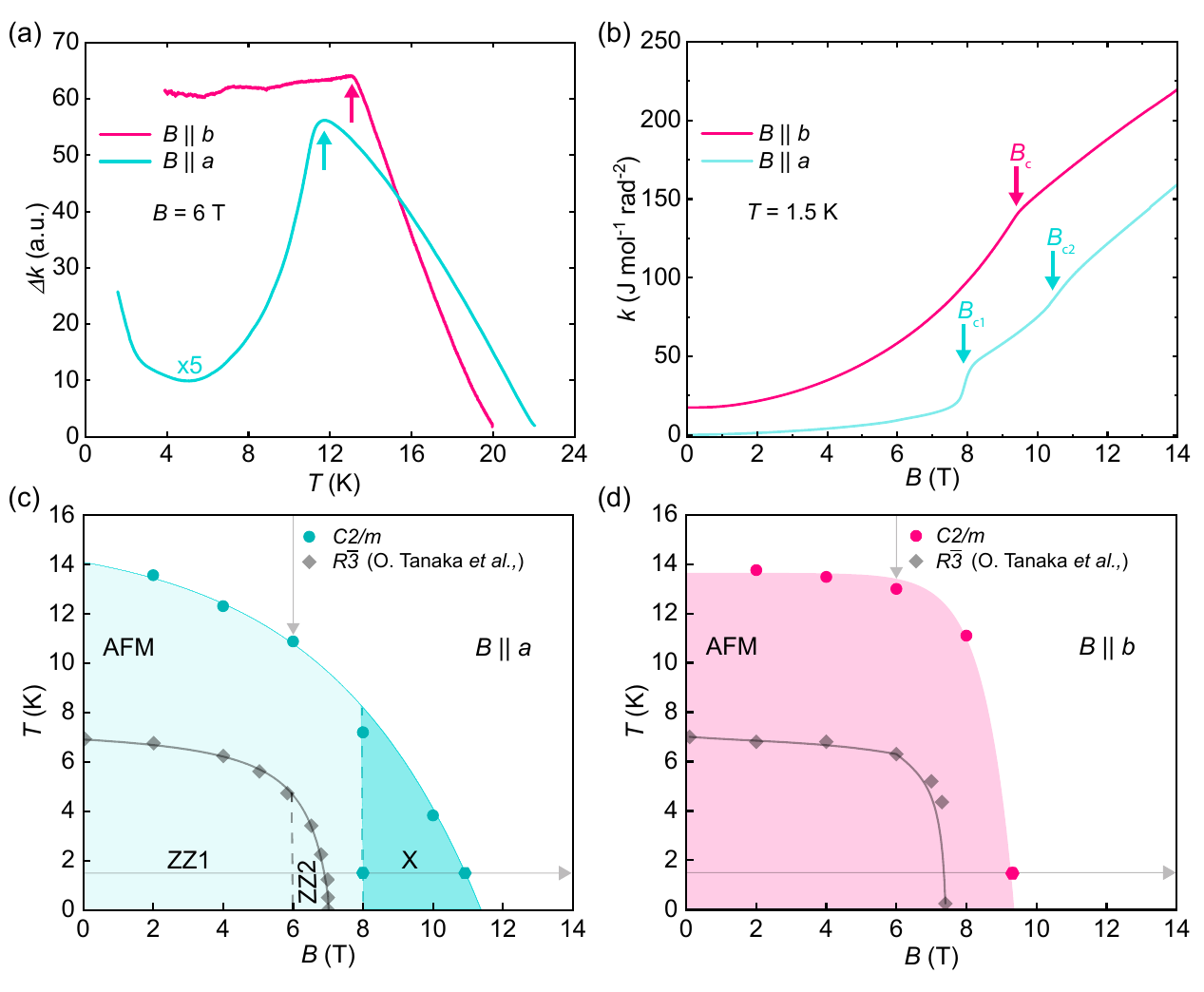}
  \caption{
  (a) Magnetotropic susceptibility as a function of temperature at  \textit{B} = 6 T, measured with the magnetic field applied along the in-plane \textit{a} and \textit{b} crystallographic axes. A pronounced kink marks the transition into the ordered state. The $a$-axis curve has been scaled for clarity to allow direct comparison with the $b$-axis data. 
  (b) Magnetotropic susceptibility as a function of magnetic field at \textit{T} = 1.5 K for fields along the \textit{a} and \textit{b} directions. The arrows represent the critical fields.
  (c-d) Temperature–field (\textit{T–B}) phase diagrams of \rucl\ for the rhombohedral \rhom phase (adapted from Ref.~\cite{Tanaka2022}, gray diamonds) and the AFM phase of the monoclinic $C2/m$ structure as determined from our measurements (cyan circles for $B||a$ and pink circles for $B||b$). The vertical dashed line marks the intermediate \textit{ZZ2} phase in the rhombohedral crystals and the corresponding \textit{X} phase in the monoclinic crystals. The AFM transition points are obtained from temperature sweeps at fixed fields from 2-10 T for $B||a$ and 2-8 T for $B||b$. Light gray vertical arrows indicate the temperature sweeps done at $B = 6$ T in panel (a) and the field sweeps at $T = 1.5$ K in panel (b).
}
  \label{fig:figure3}
\end{figure*}

\autoref{fig:figure2}(a) shows an SEM image of a small \rucl crystal (Sample 1) mounted on the end of a silicon cantilever. The sample is aligned to detect anisotropy in the $bc^\star$-plane. \autoref{fig:figure2}(b) shows a schematic temperature-field phase diagram of \rucl for $B||b$, which has been intensely investigated \cite{Balz2019,Sears2017,Balz2021,Banerjee2017a,Cao2016,Kubota2015,Baek2017,Wolter2017,Kasahara2018,Kasahara2018a,banerjee2018excitations,majumder2015anisotropic,leahy2017anomalous,zheng2017gapless,johnson2015monoclinic}. The dark pink region shows the AFM phase. The light purple region denotes the so-called Kitaev paramagnet (KPM), where the Kitaev exchange interactions ($K \approx 7$~meV $\sim 81$~K) dominate over the thermal energy scale \cite{Banerjee2017a, Do2017}. Spins in this regime are strongly correlated, but remain disordered, in contrast to the conventional high-temperature paramagnet where spins act independently (white region).


The magnetotropic susceptibility (\textit{k}) of \rucl was measured up to 14~T at 1.5~K for several in-plane field orientations (\autoref{fig:figure2}(c)). At low fields, $k$ follows a quadratic-in-field dependence that continues until the AFM phase transition $B_\text{c}$ is reached. At the phase transition, the size of the jump $\Delta k$ in the magnetotropic susceptibility depends strongly on the shape of the AFM phase boundary \cite{Modic2020}. In the case of highly-anisotropic magnets, such as \rucl, qualitative differences in the jump are expected for different field angles; $\Delta k$ will be minimal for all in-plane field angles, and enhance significantly as the magnetic field is rotated toward the out-of-plane direction due to the strong enhancement of $B_\text{c}$ for this field direction (SI \S V.1).

Despite the fact that $\Delta k$ is expected to be small for in-plane field orientations, two transitions are clearly visible for $B||a$: $B_\text{c1}$ at 7.9~T and $B_\text{c2}$ at 10.7~T (\autoref{fig:figure2}(c)). 
These transitions are intrinsic to the \mono structure, and the intervening phase between them is reminiscent of the ordered zigzag phase ($ZZ2$) reported in \rhom crystals of \rucl \cite{Sears2017, Balz2021, Wagner2022}. $B_\text{c1}$ and $B_\text{c2}$ are also highly reproducible; between the measurements taken at $\phi = 90^\circ$ and $\phi = 268^\circ$ in \autoref{fig:figure2}(c), the sample is warmed to room temperature and rotated by $\sim$180$^\circ$---the fact that no additional transitions appear confirms that no additional strain-induced phases are formed during this process. 
While there are no indications of ordered AFM phases associated with the \rhom structure in \autoref{fig:figure2}(c), two small peaks are resolvable in the field-derivative of $k$ curves for $B||a$ (SI Fig.~S5). This indicates that a very small fraction of our sample contains the \rhom structure, in agreement with our structural determination. The additional small peaks appear at magnetic fields below $B_\text{c1}$ and $B_\text{c2}$, at the same fields needed to suppress $ZZ1$ and $ZZ2$ in \rhom crystals for $B||a$ \cite{Sears2017, Balz2021, Wagner2022} (\autoref{fig:figure5_one panel}). 

As the in-plane magnetic field is rotated roughly 30$^\circ$ away from the $a$-axis, only a single transition is observed. This behavior continues until $B||b$, where the single transition occurs at $B_\text{c} \simeq 9.2$ T. No further transitions are resolvable, even in the field-derivative of $k$, for $B||b$ (SI Fig.~S5).

\autoref{fig:figure3}(c,d) shows the temperature-field ($T$-$B$) phase diagram of the \mono phase for $B||a$ and $B||b$. They are established by temperature and field sweeps of $k$, examples of which are shown in \autoref{fig:figure3}(a,b). The AFM phase boundary for the \rhom phase as determined by specific heat measurements is shown for comparison in gray \cite{Tanaka2022}. \autoref{fig:figure3}(a) shows a temperature sweep at $B = 6$ T for both field orientations. The transition temperatures differ by nearly 2~K, directly reflecting the in-plane anisotropy of the AFM phase in a magnetic field. \autoref{fig:figure3}(b) shows the field sweeps at $T= 1.5$ K (reproduced from \autoref{fig:figure2}(c)). For $B||a$ (blue curve), the first transition corresponds to entry into the intermediate phase, which we denote as $X$. The second transition $B_\text{c2}$ marks the suppression of AFM order \cite{Kim2024,Cen2025}. Note that $B_\text{c2} = 10.7$~T for $B||a$ corresponds to the same field beyond which oscillations in the thermal conductivity are no longer observed \cite{Czajka2021, Bruin2023, Lefrancois2023}. A similar field scale of approximately 11 T for $B||a$ has been associated with a proposed transition between topologically distinct quantum spin-liquid phases \cite{suetsugu2022evidence}. Several other measurements have identified an anomaly at $\sim10.7$~T, interpreting it as a crossover or a weak phase transition without resolving its origin \cite{Bruin2022,gass2020field,Tanaka2022,bachus2020thermodynamic}.

The shape of the AFM phase in the \mono crystal is qualitatively similar to that observed in \rhom samples; for $B||a$ it is more rounded, whereas for $B||b$, it remains flatter at low fields and then has a more sudden drop downwards at higher fields \cite{Tanaka2022}. While the AFM phases in the \mono and \rhom phases are qualitatively similar, the critical temperatures and magnetic fields are systematically shifted to higher values in the \mono phase. 

\begin{figure*}[t]
  \centering
  \includegraphics[width=0.8\textwidth, trim=0cm 0cm 0cm 0cm, clip=true]{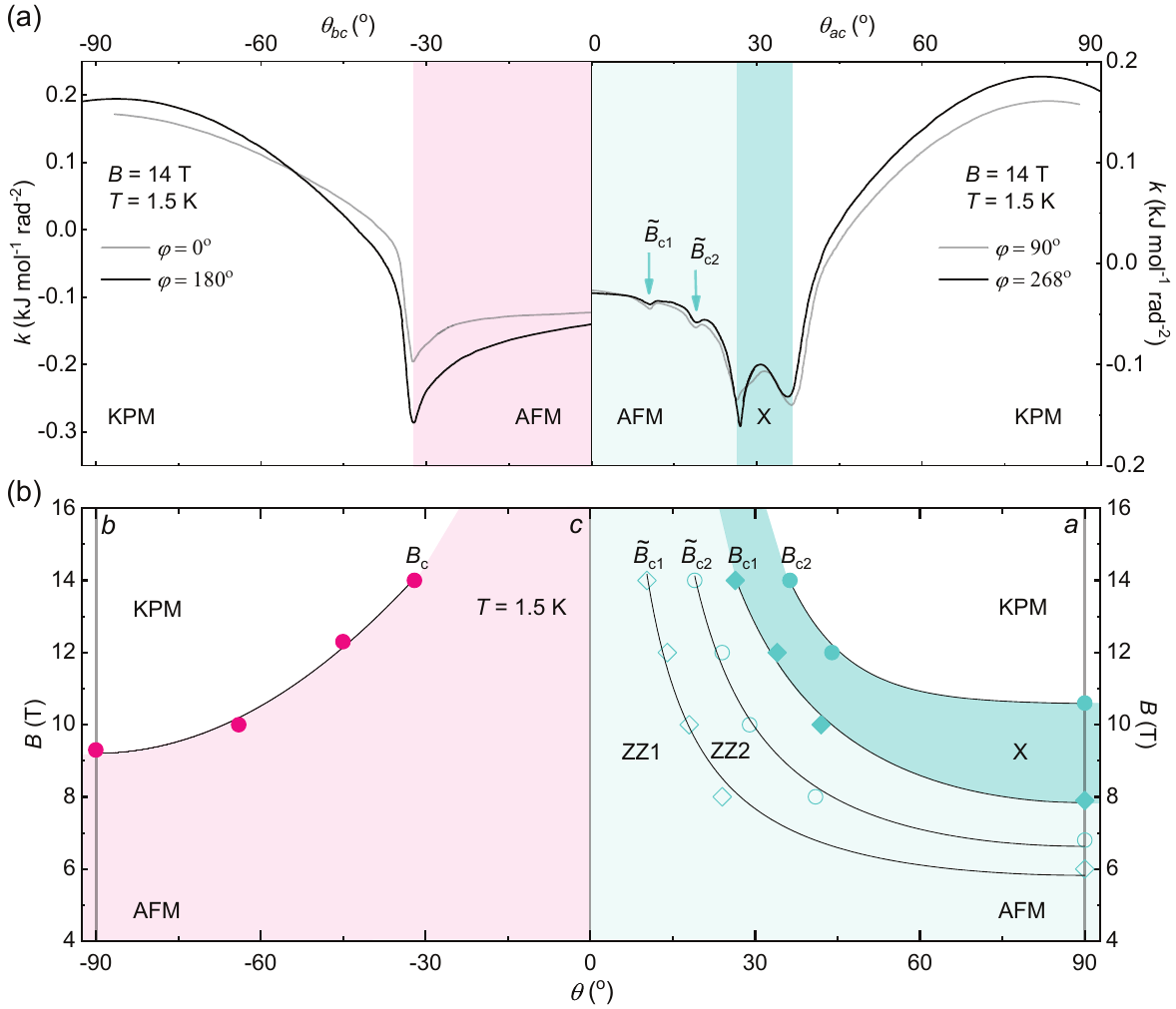} 
  \caption{
  (a) Angular dependence of the magnetotropic susceptibility $k(\theta)$ measured at $B = 14$ T and $T = 1.5$ K in the $bc$-plane (left panel, pink) and $ac$-plane (right panel). For rotation within the $bc$-plane, a single transition between the AFM and KPM states is observed. In contrast, rotation within the $ac$-plane (right panel, teal) reveals two distinct anomalies at $B_{c1}$ and $B_{c2}$, corresponding to the intermediate field-induced $X$ phase of the monoclinic structure. Two additional anomalies, labeled $\tilde{B}_{c1}$ and $\tilde{B}_{c2}$, are attributed to the $ZZ1$ and $ZZ2$ transitions of the coexistence of rhombohedral (\rhom) domains. 
  (b) Angle-dependent critical fields at $T = 1.5$ K for a nominally monoclinic \rucl crystal. The black lines represent $1/\cos\theta$ fits that extrapolate our data to the $ZZ1$ and $ZZ2$ transition fields reported for rhombohedral crystals \cite{Tanaka2022}.
  }
  \label{fig:figure4}
\end{figure*}

\begin{figure}[ht]  
  \centering
  \includegraphics[width=1\columnwidth]{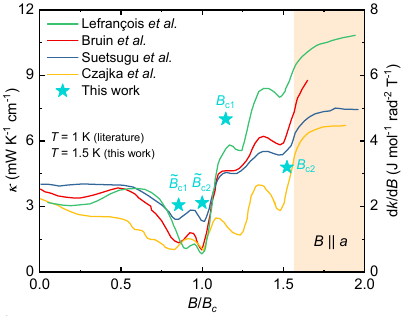}
  \caption{Comparison of magnetotropic susceptibility and thermal conductivity in \rucl for in-plane magnetic fields. Data are plotted as a function of normalized field $B/B_c$, where $B_c$ denotes the critical field at which the antiferromagnetic order of the rhombohedral phase is suppressed. Magnetotropic susceptibility measured at $T = 1.5$ K (cyan stars—the field derivative of the magnetotropic susceptibility) is compared with thermal conductivity data from four independent studies: Czajka et al. (yellow, $T = 0.96$ K) \cite{Czajka2021}, Bruin et al. (red, $T = 1.0$ K) \cite{Bruin2023}, Suetsugu et al. (blue, $T = 1.0$ K) \cite{suetsugu2022evidence}, and Lefrançois et al. (green, $T = 1.16$ K) \cite{Lefrancois2023}, each normalized by their respective $B_c$ values. In all cases, the magnetic field is applied along the $a$ axis. Four characteristic features are identified at $\tilde{B}_{c1}$, $\tilde{B}_{c2}$, $B_{c1}$, and $B_{c2}$. The two minima at higher fields ($B_{c1}$, $B_{c2}$) correspond to transitions associated with the monoclinic phase, while the smaller minima ($\tilde{B}_{c1}$, $\tilde{B}_{c2}$) are attributed to the rhombohedral component. The close alignment of these features across all datasets demonstrates a consistent set of field-induced transitions. The shaded region indicates the high-field regime where no further anomalies are resolved.
}
  \label{fig:figure5_one panel}
\end{figure}

 We now explore the angular range over which the $X$ phase exists in a single crystal with nominal \mono symmetry. Because \rucl is quasi-2D, rotating the magnetic field from the $c^\star$-axis into the $ab$-plane is, to first approximation, equivalent to an in-plane magnetic field sweep \cite{Modic2020}. \autoref{fig:figure4}(a) shows the angle dependence of $k$ measured at $B = 14$ T and $T = 1.5$ K. When the sample is rotated through the planes at $\phi = 0^\circ$ and $\phi=180^\circ$ (\textit{i.e.} the $bc$-plane in the left panel) in a large fixed magnetic field, the sharp minimum in $k(\theta)$ coincides with the AFM phase boundary. The larger out-of-plane field component produces a larger thermodynamic signature due to the steep curvature of the AFM phase boundary as magnetic field is rotated perpendicular to the honeycomb planes (SI \S V.1). Only one transition, $B_\text{c1}$, from AFM order into the Kitaev paramagnet, is observed at $\theta \approx -35^\circ$, consistent with the field sweep along the $b$-axis (\autoref{fig:figure3}(b)). For field rotation in the $bc$-plane at lower fixed magnetic fields, the critical fields are observed at larger $|\theta|$, which is summarized in the left panel of \autoref{fig:figure4}(b). This behavior is consistent with the fact that the lowest critical fields in \rucl are observed for in-plane field orientations \cite{Modic2018, Modic2020,Zhou2023,Gordon2019a,riedl2019sawtooth,li2021identification,wang2019one}.

When a large magnetic field ($\geq$8 T) is rotated within the $ac$-plane, four transitions are observed as sharp minima in $k$. The right side of \autoref{fig:figure4}(a) shows the angle dependence of $k$ measured at $B = 14$ T and $T = 1.5$ K when the sample is rotated through the planes at $\phi = 90^\circ$ and $\phi=268^\circ$ (\textit{i.e.} the $ac$-plane). The two larger transitions (at $\theta = 27^\circ$ and $\theta = 36^\circ$) correspond to $B_\text{c1}$ and $B_\text{c2}$, which delineate the $X$ phase of the \mono structure. The other two additional features at lower angles (closer to the $c^\star$-axis), which we label as $\tilde{B}_{\text{c1}}$ and $\tilde{B}_{\text{c2}}$, extrapolate to the $ZZ1$ and $ZZ2$ AFM transitions observed under in-plane magnetic fields ($\theta = 90^\circ$) in \rhom crystals \cite{Tanaka2022,Balz2019,Balz2021} (right side of \autoref{fig:figure4}(b)).


\autoref{fig:figure5_one panel} (as well as SI Fig.~S11(a)) shows our measurements in the context of previous reports, first assembled by \citet{Lefrancois2023}. The critical fields we find (cyan stars) overlaid with their thermal transport data show that the minima of the oscillatory thermal response agree with the magnetic phase transitions identified here \cite{Lefrancois2023,Bruin2023,suetsugu2022evidence,Czajka2021}. A key insight made by \citet{Lefrancois2023} was that the same number of oscillations occurred regardless of the in-plane field orientation even though the field window of the oscillations changed by roughly a factor of 2, leading to the conclusion that the oscillations are not tied to a fixed Fermi surface geometry and therefore do not support the existence of a spinon Fermi surface.

Our observations directly link the thermal anomalies to the magnetic phase diagram, providing compelling evidence that the reported features arise from multiple magnetic transitions associated with coexisting \rhom and \mono structural domains. This picture is also consistent with a recent study using ultraclean crystals, which found that the oscillatory anomalies persist but are significantly attenuated relative to conventionally grown crystals \cite{Xing2025}. Such behavior is naturally expected if the anomalies scale with the residual fraction of coexisting structural domains, as we establish here (SI \S V.2). 


This comparison extends well beyond thermal transport anomalies---Table~\ref{tab:tn_bc_compilation} shows an extended body of literature reporting multiple critical fields under in-plane magnetic fields in \rucl. Across diverse measurement methods,  a majority of those critical fields coincide with the $\tilde{B}
_\text{c}$ transitions of the \rhom phase and the $ B
_\text{c}$ transitions of the \mono phase. Furthermore, studies where fewer field-induced transitions are resolved tend to correlate with a dominant $T_\text{N}$, indicating that the sample is either structurally-pure or sufficiently small that it remains predominantly \mono.

\section{Discussion}

The combination of structural characterization and magnetotropic susceptibility measurements establishes the intrinsic magnetic phase diagram of monoclinic ($C2/m$) RuCl$_3$. We find that the $T$--$B$ phase diagram of the $C2/m$ structure closely resembles that previously reported for the rhombohedral ($R\bar{3}$) phase, but with systematically enhanced energy scales. In our crystal, the AFM transition occurs at a single transition $T_\text{N} = 14$ K, and the critical field reaches $B_{c} \approx 9.2$ T for $B||b$, in agreement with earlier torque measurements on small crystals \cite{Modic2018,Modic2020}. In addition, for magnetic fields applied within $\lesssim 30^\circ$ of the $ac$-plane, we resolve the boundary of what appears to be a second, long-range magnetically-ordered state $X$---analogous to the field-induced $ZZ2$ phase observed in $R\bar{3}$ samples \cite{Balz2021,Tanaka2022, bachus2020thermodynamic,bachus2021angle}. The transitions surrounding the $X$ phase are most apparent in the angle dependence, where $B_{c1}(\theta)$ and $B_{c2}(\theta)$ trace non-merging phase boundaries to the zero-field, zero-temperature limit (\autoref{fig:figure4}b). Future work, such as resonant x-ray scattering, will be required to determine the magnetic texture of the $X$ phase on relatively small, \mono samples. 

\begin{table*}
\begin{threeparttable}
\caption{Compilation of N\'{e}el temperatures and critical fields for $\alpha$-RuCl$_3$ across the literature, aligned to the four $B\parallel a$ critical fields and one $B\parallel b$ critical field of this work. \textbf{Bold} values are the main/dominant feature reported in that dataset; unbolded values are weak/secondary features. Where a critical field was not explicitly stated in the source text, its value was estimated by visual inspection from the data shown in the published figures. Values are taken from the curve closest to $T=1.5$~K in each study. Sample mass is an order-of-magnitude estimate. Orientation: C = confirmed, field direction explicitly stated in the source; I = inferred, not stated and placed by closest numerical match to this work's values. TMR = tunneling magnetoresistance; MCE = magnetocaloric effect.}
\label{tab:tn_bc_compilation}
\small
\setlength{\tabcolsep}{8pt}
\begin{tabular}{l@{\hspace{4pt}}l@{\hspace{4pt}}c@{\hspace{4pt}}c cccccl}
\toprule
Source & Technique & Sample Mass & $T_N$ (K) & \multicolumn{4}{c}{$B\parallel a$ (T)} & $B\parallel b$ (T) & Or. \\
\midrule
\textbf{This work} & RTM & 10~ng & \textbf{14} & 5.8 & 6.9 & \textbf{7.9} & \textbf{10.7} & \textbf{9.2} & C \\
\midrule
Czajka 2021 \cite{Czajka2021} & $\chi$, $\kappa_{xx}$ & 1~mg & \textbf{7.8} & \textbf{5.9} & \textbf{7.3} & 8.9 & \textbf{10.7} & -- & C \\
Lefran\c{c}ois 2023 \cite{Lefrancois2023} & $\kappa_{xx}$ & 1~mg & \textbf{7} & \textbf{6.9} & 7.8 & 9.2 & \textbf{11.1} & -- & C \\
Bruin 2022 \cite{Bruin2023} (\textit{Origin}) & $\kappa_{xx}$ & 100~$\mu$g & \textbf{7.5}, 10, 14 & \textbf{5.9} & \textbf{7} & 8.5 & 10.5 & \textbf{9}\tnote{a} & C \\
Bruin 2022 \cite{Bruin2022} (\textit{Robust}) & $\kappa_{xx}$ & 100~$\mu$g & 7.5 & 6 & 7.2 & 8.7 & 10.7 & -- & I \\
Suetsugu 2022 \cite{suetsugu2022evidence} & $\kappa_{xx}$ & -- & \textbf{7}, 14 & 6 & 7.2 & 9.5 & 10.7 & \textbf{7.6}\tnote{b} & C \\
Balz 2019 \cite{Balz2019} & MCE & 1~g & \textbf{7} & 6.5 & \textbf{7.3} & 9.1 & 10.2 & -- & I \\
Sears 2017 \cite{Sears2017} & $C$, neutron & 100~mg & \textbf{6.7}, 9.2 & -- & -- & 8.1 & -- & -- & I \\
Kubota 2015 \cite{Kubota2015} & $\chi$ & -- & \textbf{7}, 14 & 6 & 7.5 & 8 & 9.5 & -- & I \\
Kubota 2015 \cite{Kubota2015} & $C$ & -- & \textbf{7.8}, 10, 12, 14 & -- & -- & 7.9 & -- & -- & I \\
Tanaka 2022 \cite{Tanaka2022} & $C$ & -- & -- & \textbf{5.8} & \textbf{6.9} & -- & 10 & 7 & C \\
Balz 2021 \cite{Balz2021} & $\chi$ & 1~g & 6 & 6.1 & \textbf{7.1} & -- & -- & \textbf{7.6} & C \\
Balz 2021 \cite{Balz2021} & $\kappa_{xx}$ & 1~g & -- & \textbf{6.9} & 7.8 & -- & -- & -- & C \\
Xing 2025 \cite{Xing2025} & $\kappa_{xx}$ & 10~mg & \textbf{7.8} & \textbf{7.1} & \textbf{8.2} & 9 & 9.8 & \textbf{8.7}\tnote{c} & C \\
Massicotte 2024 \cite{massicotte2024ACS} & TMR & few layers & \textbf{14} & -- & -- & -- & \textbf{10.7} & \textbf{9.4} & C \\
\bottomrule
\end{tabular}
\begin{tablenotes}
\small
\item[a] Bruin 2022 (\textit{Origin}) $B\parallel b$ shows four transitions: 7, \textbf{7.5}, 8.5, \textbf{9}~T (9~T shown as nearest match).
\item[b] Suetsugu 2022 $B\parallel b$ shows four transitions: 6, \textbf{7.6}, 9, 10.5~T (7.6~T shown as nearest match).
\item[c] Xing 2025 $B\parallel b$ shows two transitions: 8.1, \textbf{8.7}~T (8.7~T shown as the main/dominant feature).
\end{tablenotes}
\end{threeparttable}
\end{table*}

The similarities between the phase diagrams are surprising given the structural differences between the two polymorphs. When RuCl$_3$ transforms from $C2/m$ stacking at high temperature to $R\bar{3}$ stacking at low temperature, the in-plane Ru--Ru bond lengths become equal, restoring global threefold symmetry \cite{Kim2024}. In the $C2/m$ structure, by contrast, the three honeycomb bonds are no longer symmetry-equivalent. This permits more symmetry-allowed ligand displacements and different octahedral tilts on different neighboring bonds, modifying the Ru--Cl--Ru exchange pathways. Given how sensitively the Kitaev and $\Gamma$ interactions depend on the exchange pathways mediated by the ligand ions \cite{chaloupka2013zigzag}, there is no reason to believe that the magnetic phase diagrams of $C2/m$ and $R\bar{3}$ symmetry should be the same, even qualitatively. Yet, our results show that the AFM phases for both structures have the same qualitative shape, and they both enclose a second AFM phase near $B||a$.

The primary distinction between the two phase diagrams is quantitative. In the $C2/m$ structure, the characteristic temperatures and fields are uniformly enhanced. This suggests that the unequal bond lengths primarily reduce magnetic frustration without fundamentally altering the hierarchy of exchange interactions that govern the magnetism. One way to achieve the same qualitative shape of the AFM phases of both structures is if the dominant couplings ($J$, $K$, $\Gamma$, and further-neighbor Heisenberg terms) scale together (i.e.\ without changing their relative magnitudes or signs). Such robustness has important implications for attempts to tune the system toward a spin-liquid regime. If moderate structural distortions simply renormalize the overall energy scale, then pressure or strain may shift phase boundaries without dramatically altering the underlying magnetic Hamiltonian.

Recent theoretical and experimental studies have indeed highlighted the strong magnetoelastic coupling in $\alpha$-RuCl$_3$, suggesting that uniaxial strain or pressure can significantly modify exchange interactions \cite{Wolf2022, Bastien2018}. Our results motivate revisiting such approaches in crystals that remain purely $C2/m$ at low temperature, where structural inhomogeneity is minimized.

Upon establishing the intrinsic phase diagram of the $C2/m$ structure, each anomaly in this work can be assigned to a well-defined thermodynamic phase transition: two corresponding to the $C2/m$ phase ($B_{c1}$ and $B_{c2}$) and two to the residual $R\bar{3}$ fraction ($\tilde{B}_{c1}$ and $\tilde{B}_{c2}$ for entering and exiting the $ZZ2$ phase). All four features evolve with a systematic angular dependence expected from the magnetic phases of both polymorphs. Our data provide a natural, thermodynamically-grounded explanation for nearly all of the field-induced phase transitions that have been routinely observed in \rucl, in the regime that’s been widely described as a field-induced Kitaev spin liquid.

The clarity gathered here further explains why experiments performed on crystals containing mixtures of $C2/m$ and $R\bar{3}$ stacking often report oscillations over a broader range of field directions than expected; measurements effectively average over domains with different stacking orientations, bringing multiple phase boundaries into the same field window. Because the intermediate magnetic phase occupies a narrow, well-defined angular window tied to the crystallographic $a$-axis, a crystal containing multiple rotated domains would show its signatures spread across a correspondingly broader range of field angles than a single, untwinned crystal. Coexisting $C2/m$ domains rotated by $120^\circ$ \cite{Cao2016, Sears2023}, together with $R\bar{3}$ domains whose principal axes are rotated relative to those of $C2/m$, would each contribute their own angular window, and orientational disorder associated with stacking faults would further broaden the field-angle space over which the AFM transitions of each structure are observed. 



Beyond resolving the origin of the oscillatory transport anomalies, the phase diagram established here provides a general reference against which field-induced measurements throughout the $\alpha$-RuCl$_3$ literature can be interpreted. Table~\ref{tab:tn_bc_compilation} compiles N\'{e}el temperatures and critical fields reported across twelve independent studies, spanning six distinct experimental techniques (thermal conductivity, magnetic susceptibility, neutron diffraction, magnetocaloric effect, specific heat, and tunneling magnetoresistance). The overwhelming majority of these values coincide with one of the four magnetic phase boundaries established here for the $C2/m$ and $R\bar{3}$ structures, despite having been measured on different crystals, by different groups, using different methods. This is a remarkable convergence given the large magnetic anisotropy of \rucl, which would readily shift these critical fields under even modest sample misalignment. That two to four of these critical fields are routinely resolved across such diverse studies, consistently landing at closely-matching values, indicates that the intrinsic magnetic phase diagram---rather than sample-specific disorder---is the organizing principle behind the great majority of field-induced signatures reported in this material. Moreover, which subset of the four transitions a given study resolves offers a structural diagnostic in its own right: studies reporting predominantly the lower pair are consistent with $R\bar{3}$-dominated crystals, those reporting predominantly the upper pair with $C2/m$-dominated crystals, and those resolving all four with genuinely mixed-phase samples. This allows the structural composition of a crystal to be inferred directly from its thermodynamic response---valuable precisely because independent, low-temperature structural characterization is often not feasible.

These results also challenge the common perception that monoclinic crystals represent lower-quality realizations of RuCl$_3$. Instead, we find that the monoclinic structure hosts essentially the same microscopic exchange interactions as the rhombohedral phase, with the principal difference being a uniform enhancement of the magnetic energy scales. By establishing the intrinsic AFM phase diagram of a structurally-pure monoclinic crystal, our work provides a clear thermodynamic reference against which previously-reported anomalies can be understood, and a solid foundation for interpreting new field-induced phenomena. Future experiments that deliberately control distortions---through strain engineering, pressure, or the synthesis of small single-domain crystals---may offer a promising route to identify the conditions under which a Kitaev spin liquid may emerge, and ultimately tune materials there.

\section*{Acknowledgments}
The authors thank Dr. Pradeep Mandal for the assistance with the SC-XRD data collection. This work was supported by the Scientific Service Units of the Institute of Science and Technology Austria (ISTA) through resources provided by the Lab Support Facility. The authors also thank the Nanofabrication Facility, another Scientific Service Unit at ISTA; in particular, we appreciate support from Evgeniia Volobueva for providing assistance and access to the PFIB and Lubuna Shafeek for assistance at the VSM. H.N. and K.A.M. thank Michael Baenitz and Marcus Schmidt of the Max Planck Institute for Chemical Physics of Solids, Dresden, for providing the \ar crystals used in our reproducibility measurements. H.N. and K.A.M. appreciate scientific discussions with B. J. Ramshaw, Shiva Safari, Arkady Shekhter, Roser Valenti, and Valeska Zambra. H.N. and K.A.M. appreciate the technical support provided by Zolt\'an K\"oll\"o.

H.N., M.N., V.N. and K.A.M. acknowledge funding received from the European Research Council (ERC) under the European Union's Horizon 2020 research and innovation programme (TROPIC-101078696) and collaborative interactions through the research project funded by the Austrian Science Fund (FWF) SFB Q-M\&S F 86. H.N. acknowledges financial support within the framework of the FTI Strategy Niederösterreich 2027 (“Gefördert im Rahmen der FTI-Strategie Niederösterreich 2027, FTI Dissertationen, Nr. FTI23-D-043.”). E.H., S.K., and Y.J.K.'s work at the University of Toronto was supported by the Natural Sciences and
Engineering Research Council (NSERC) of Canada through the Discovery Grant No. RGPIN-2025-06514, and by the Canada Foundation for Innovation (CFI) and the Government of Ontario for Project No. 36404.

The content does not necessarily represent the perspective of the federal state of Lower Austria or the Gesellschaft für Forschungsförderung Niederösterreich as the funding agency. Neither the federal state of Lower Austria nor the funding agency can therefore be held responsible for the content.

\section*{Author contributions}
H.N. and K.A.M. conceived of the experiment; E.H., S.K., and Y.J.K. provided the samples; H.N. performed the experiments and analyzed the data, with support from D.B., M.N., and V.N.; H.N. and K.A.M. wrote the manuscript with input from all co-authors.


\setcounter{NAT@ctr}{0}
\putbib[RuCl3_refs,RuCl3_refs_2]

\clearpage
\end{bibunit}


\clearpage

\onecolumngrid

\begin{bibunit}[apsrev4-2]

\onehalfspacing
\setlength{\parskip}{1em}
\setlength{\parindent}{0pt}

\renewcommand{\kim}[1]{{\color{Red} #1}}
\renewcommand{\mono}{$C2/m$\xspace}
\renewcommand{\rhom}{$R\bar{3}$\xspace}

\pagenumbering{arabic} 

\setcounter{section}{0}
\setcounter{subsection}{0}
\setcounter{subsubsection}{0}
\setcounter{secnumdepth}{3}

\setcounter{figure}{0}
\renewcommand{\thefigure}{S\arabic{figure}}

\setcounter{table}{0}
\setcounter{equation}{0}



\section*{Supplementary information for ``A thermodynamic reference for field-induced phenomena in \ar: two coexisting polymorphs account for multiple magnetic phase transitions''}


\section{Structural analysis}
\label{sec:xray}
\subsection{Identifying the monoclinic phase via reciprocal-space mapping}

Reciprocal-space maps (RSMs) for Sample S2 were analyzed using an orthorhombic (rectilinear) coordinate system to facilitate direct comparison between monoclinic (\mono) and rhombohedral (\rhom) structural descriptions reported in the literature \cite{Kim2024}. The diffraction data were first indexed in the monoclinic \mono structure established in Ref.~\cite{johnson2015monoclinic}, which uniquely determines the lattice parameters and crystal orientation. Following the procedure introduced by Kim et al. \cite{Kim2024}, the monoclinic lattice was then orthogonalized without changing the unit-cell volume. In this construction, the in-plane $a$- and $b$-directions are preserved, while the rectilinear $c$-axis is defined along the reciprocal-lattice direction $c^\star$. This representation suppresses the small monoclinic distortions and provides a convenient reference frame for identifying stacking-related structural signatures. Miller indices $h$, $k$, and $l$ therefore refer to this rectilinear basis, and diffraction peaks associated with a different crystal structure will appear at non-integer $(hkl)$ coordinates in this representation.

\begin{figure*}[h!]
  \centering
\includegraphics[width=0.9\textwidth, trim = 0cm 3cm 0cm 3cm, clip=true ]{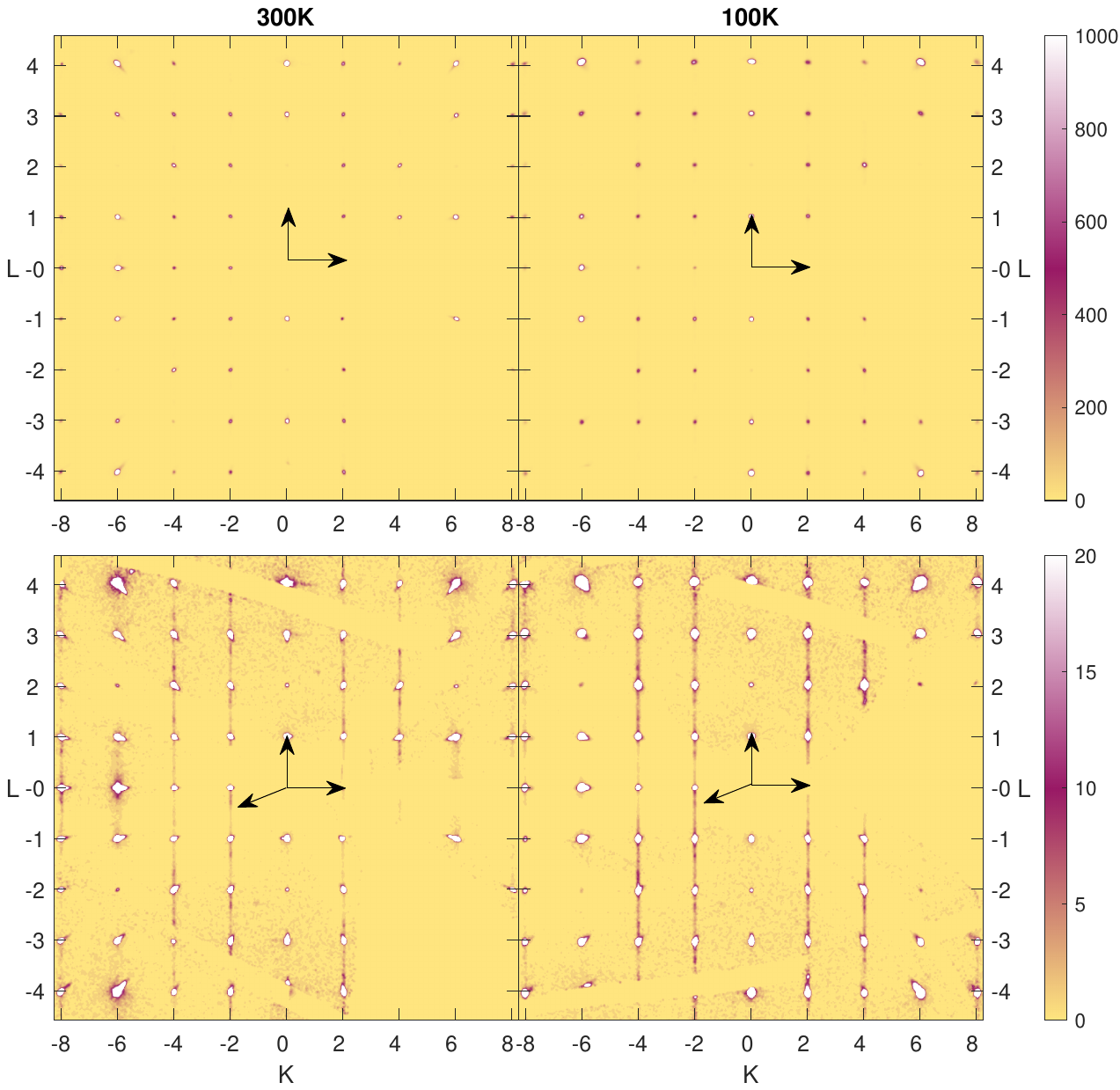} \caption{\textbf{ Reciprocal space maps (RSMs) of the $(0kl)$ plane of RuCl$_3$ at 300~K (left) and 100~K (right).} RSMs are shown on a standard intensity scale (top), exhibiting a rectangular grid of integer-index Bragg reflections (grid marked by the perpendicular arrows on the top panels) consistent with a well-ordered monoclinic structure. The same data are shown on an intensity scale enhanced to highlight weak features (bottom). Additional reflections appearing at $n \pm 1/3$ non-integer-index positions (marked by the third arrow on the bottom panels) stem from twin domains.}
\label{fig:okl}
\end{figure*}

RSMs in the $(0kl)$ plane at 300 K, shown in the top left panel of \autoref{fig:okl}, exhibit a rectangular grid of integer-index Bragg reflections, consistent with a well-ordered monoclinic structure. Within this coordinate system, the $(0kl)$ plane directly probes the relationship between the in-plane armchair periodicity (encoded in \textit{k}) and the out-of-plane stacking periodicity (encoded in \textit{l}). As discussed in Ref.~\cite{Kim2024}, rhombohedral stacking produces characteristic Bragg peaks at positions $l=n+k/6$ (with integer $n$), reflecting a modified stacking periodicity; such features are not observed here, nor at 100 K (top right), as expected for ideal monoclinic stacking. When the map is plotted to enhance weak intensity (bottom panels), additional reflections at $n \pm 1/3$ become visible, forming an oblique lattice. These peaks cannot be indexed within the primary monoclinic lattice. While their positions resemble those expected for a rhombohedral phase, we show in Section~I.2 that they instead arise from a small twin domain rotated by $120^\circ$ around $c^\star$.


\subsection{Twinning versus rhombohedral inclusions}
\label{sec:xray2}
\begin{figure*}[h]
  \centering
  \includegraphics[width=1\textwidth, clip=true ]{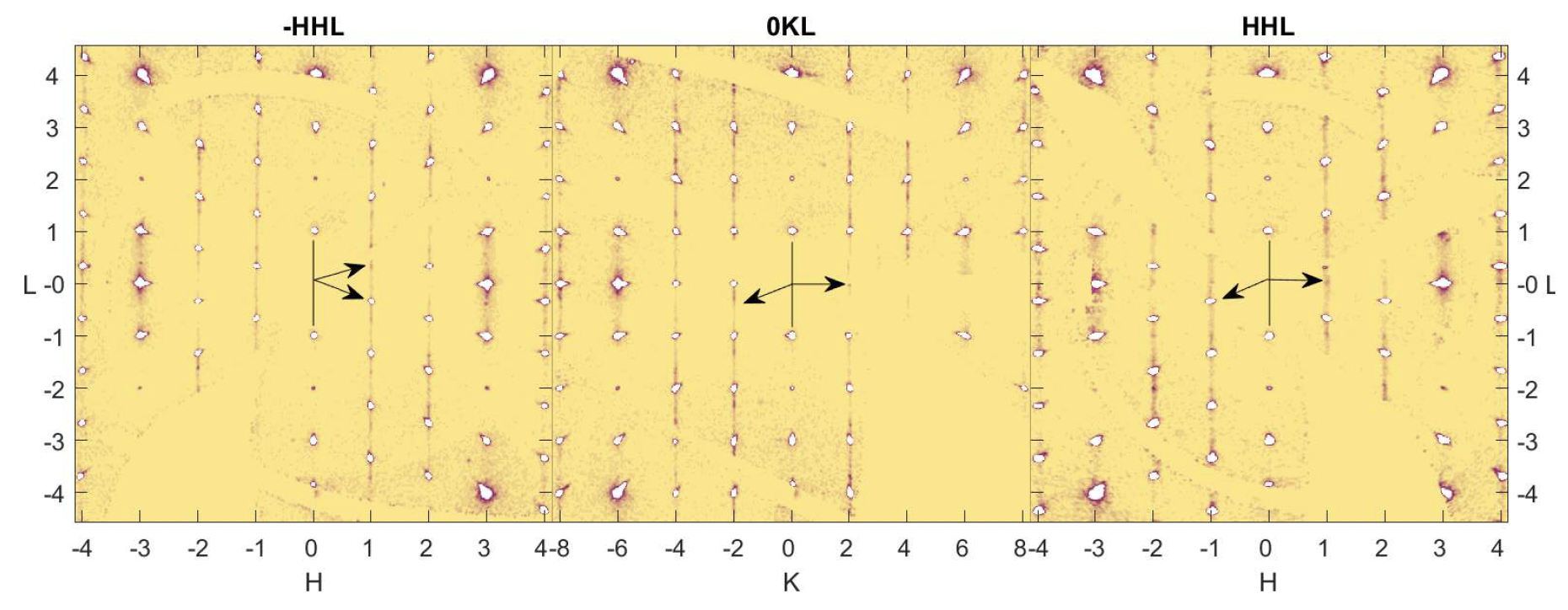}
  \caption{\textbf{Comparison of the $(-hhl)$, $(0kl)$, and $(hhl)$ reciprocal-space planes of \rucl at 300~K, distinguishing twin domains from a rhombohedral phase.} RSM of the $(hhl)$ plane, approximately corresponding to the twin domain's own $(0k'l')$ plane. The weak, fractionally-indexed reflections identified in the $(0kl)$ plane (middle) reorganize here into a rectangular lattice of integer-index peaks, while the reflections that are strong in $(0kl)$ instead appear at fractional positions $l \sim n + h/3$. This reversal of which reflection set is strong versus weak under a change of reference frame is the defining signature of a $120^\circ$-rotated twin domain of the same $C2/m$ structure, rather than a structurally distinct rhombohedral phase, whose characteristic fractional positions ($l = n + k/6$) would instead persist in all three projections regardless of viewing geometry. The RSM of the $(-hhl)$ plane, showing two misoriented twin variants; the absence of a third confirms that only two of the three possible $120^\circ$-rotated $C2/m$ twin orientations are present in this sample.}
 \label{fig:twins}
\end{figure*}

When the $(0kl)$ map is plotted to enhance intensity, additional features are observed at non-integer values (bottom panels in \autoref{fig:okl}). To determine whether these features originate from a genuine rhombohedral inclusion, we compare RSMs in the $(hhl)$, $(0kl)$, and $(-hhl)$ planes at 300~K, shown in \autoref{fig:twins}. The $(hhl)$ plane corresponds approximately to the twin domain's own $(0k'l')$ plane, where primed indices denote Miller indices expressed in the twin's rotated reference frame. In this projection, the weak, fractionally-indexed reflections identified in the $(0kl)$ plane reorganize into a rectangular lattice of integer-index peaks, while the reflections that are strong in $(0kl)$ instead appear at fractional positions $l \sim n + h/3$. This reversal of which reflection set is strong versus weak, under a simple change of reference frame, is the defining signature of a $120^\circ$-rotated twin domain of the \emph{same} $C2/m$ structure: the apparent fractional indices arise from misalignment between two domains of one structure, not from an intrinsic change in stacking periodicity. By contrast, a genuinely rhombohedral phase would produce systematic fractional peak positions ($l = n + k/6$) in all three projections, independent of viewing geometry, rather than reorganizing into an integer lattice under this specific rotation.

The third projection, the $(-hhl)$ plane, shows two misoriented twin variants, confirming that only two of the three possible $120^\circ$-rotated $C2/m$ twin orientations are present in this sample. The diffraction pattern is thus dominated by the high-quality monoclinic phase, with only a minor contribution from twinned domains. The same set of RMS maps appears remarkably similar at 100 K; the same dominant monoclinic and minority twin phases are present. We note that increased diffuse scattering is observed along the $L$ direction at 100K, indicating stacking faults; this diffuse background could mask a weak rhombohedral signal coexisting with the dominant twin contribution, so the data do not rule out a small $R\bar{3}$ fraction entirely.

\section{Vibrating sample magnetometry (VSM)}

To further assess the size-dependent effects, vibrating sample magnetometry (VSM) measurements were performed on a significantly larger crystal (S3) from the same growth batch. The volume of Sample S3 is approximately $6 \times 10^{4}$ times greater than those used for the RTM (S1) and SC-XRD (S2) measurements. While no clear signatures of a structural transition were resolvable in the smaller samples, Sample S3 exhibits a region of hysteresis that is consistent with a structural transition from the high-temperature monoclinic phase to the low-temperature rhombohedral phase.

The observation of this transition in the larger crystal confirms that the absence of a clear structural signature in the smaller samples is not due to poor sample quality. Previous studies have shown that structural disorder can give rise to multiple transitions and elevated $T_N$, and such effects are common in van der Waals materials, particularly in \rucl, leading to variability across samples \cite{Kubota2015, Mi, johnson2015monoclinic, Zhang2024, Cao2016}. The well-defined structural transition and the Néel temperature $T_N \approx 7.5$~K observed for Sample S3 indicate the high crystalline quality of the batch.

\begin{figure}[ht]
  \centering
  \includegraphics[width=0.75\columnwidth, clip=true ]{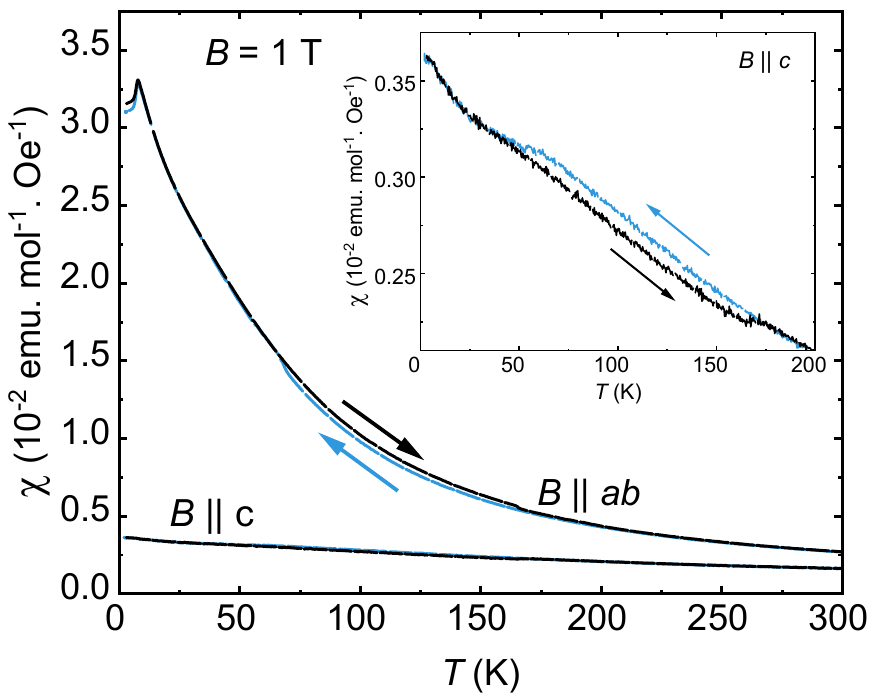}
  \caption{\textbf{Temperature-dependent magnetic susceptibility of a bulk \rucl crystal (S3).} Magnetic susceptibility $\chi(T)$ measured at $B = 1$~T for fields applied parallel to the honeycomb plane ($B \parallel ab$) and perpendicular to the plane ($B \parallel c$). Arrows indicate the cooling and warming directions. A clear hysteresis is observed for $B \parallel ab$, consistent with a first-order structural transition. Inset: corresponding data for $B \parallel c$, showing a significant hysteretic response.}
 \label{fig:vsm}
\end{figure}

\section{Experimental setup of resonant torsion magnetometry (RTM)}
Magnetotropic susceptibility ($k$) measurements were carried out on a single crystal of \rucl using resonant torsion magnetometry \cite{Modic2018, Shekhter_2023}. The crystal, with lateral dimensions of approximately $70 \times 55 \times 3~\mu$m$^3$, was carefully transferred onto the tip of a silicon cantilever (Akiyama A-probe) \cite{Akiyama2010, kort2022utilization}.

All measurements were carried out in a 14~T Oxford Teslatron superconducting magnet. Magnetic field sweeps were performed from 0 to 14~T, while temperature sweeps were conducted between 20~K and 1.5~K at various fixed fields. The sample must experience thermal cycling back to room temperature between each $\phi$ orientation of the sample on the cantilever. $\phi$ denotes the rotation angle about $c^\star$, with $\phi = 0$ corresponding to the $b$-axis aligned with the long axis of the cantilever. In this work, $c^\star$ is always aligned with the normal to the surface of the cantilever. Once the sample is cooled to base temperature, the magnetotropic susceptibility is measured as a function of $\theta$---the angle between the crystallographic \cstar axis and the external magnetic field. To change $\theta$, we use an Attocube ANR240 nanopositioner to rotate the equilibrium position of the cantilever with respect to the external field in the same plane as the fundamental vibration (flapping) mode of the cantilever. 

The same experimental setup described in \citet{Modic2020} was used here; a Zurich Instruments mid-frequency lock-in (MFLI) amplifier was used to track the cantilever's resonant frequency with the phase-locked loop option. This enables sensitive detection of the frequency shift as a function of magnetic field, temperature, and field orientation.

\section{Samples}
\label{sec:samples}
In addition to Samples 1 and 2 presented in the main text, three other samples were measured in this work.
Sample S3---a substantially larger, mg-sized sample---was used for vibrating sample magnetometry (VSM) (SI Fig.~S3). To confirm reproducibility, we repeated measurements of the magnetotropic susceptibility on a second microcrystal, Sample S4, which was grown by an independent group. Finally, to extend magnetotropic measurements to a larger, predominantly \rhom sample, we measured Sample Q1-\rhom, which was from the same growth batch as Samples 1-3. A summary of the samples used in this study and their corresponding dimensions, volumes, and masses is provided in Table~\ref{tableS1}.

Samples S1, S2, S3 and Q1 in the table below were provided by the group of Dr. Young-June Kim at the University of Toronto. Each of these single crystals was obtained from the same growth batch. Sample S4 was provided by Michael Baenitz and Marcus Schmidt from the Max Planck Institute for the Chemical Physics of Solids in Dresden.

\begin{table}[h]
\centering
\setlength{\tabcolsep}{12pt}
\begin{tabular}{cccccc}
\toprule
Sample & Dimensions ($\mu$m) & Volume ($\mu$m$^{3}$) & Mass (ng) & Technique\\
\midrule
S1 & 70 x 55 x 3 & 11,550 & 35.9 & RTM\\
S2 & 50 x 60 x 4 & 12,000 & 37.2 & SC-XRD\\
S3 & 1570 x 990 x 460 & 7.15 × $10^8$ & 2.22 x $10^6$ & VSM\\
S4 (MPI) & 3849.5 $\mu$m x 2 & 7,700 & 28 & RTM\\
Q1-\rhom & 728 x 200 x 70 & 10.2 x $10^6$ & 3.7 x $10^4$ & RTM (qPlus sensor)\\

\bottomrule
\end{tabular}
\caption{Details related to the samples measured in this study. 
}
\label{tableS1}
\end{table}

Magnetotropic susceptibility measurements were performed on a micron-sized crystal (S1) using resonant torsion magnetometry (RTM).
To assess its structural properties, a separate sample (S2) of comparable size was characterized using single-crystal X-ray diffraction (SC-XRD). As described in \autoref{sec:xray2}, our X-ray data show that Sample S2 has \mono symmetry with few monoclinic twin domains. To test whether the structural symmetry depends on the sample size \cite{harford2026sample}, we selected a much larger crystal (S3) from the same growth batch and searched for the structural transition using vibrating sample magnetometry (VSM) measurements. \autoref{fig:vsm} shows a hysteretic structural transition at $\sim$120 K, confirming that the larger crystal undergoes the structural transformation to the \rhom phase. 

\begin{figure*}[ht]
\centering
\includegraphics[width=1\textwidth, trim=0cm 4.1cm 0cm 3cm, clip=true]{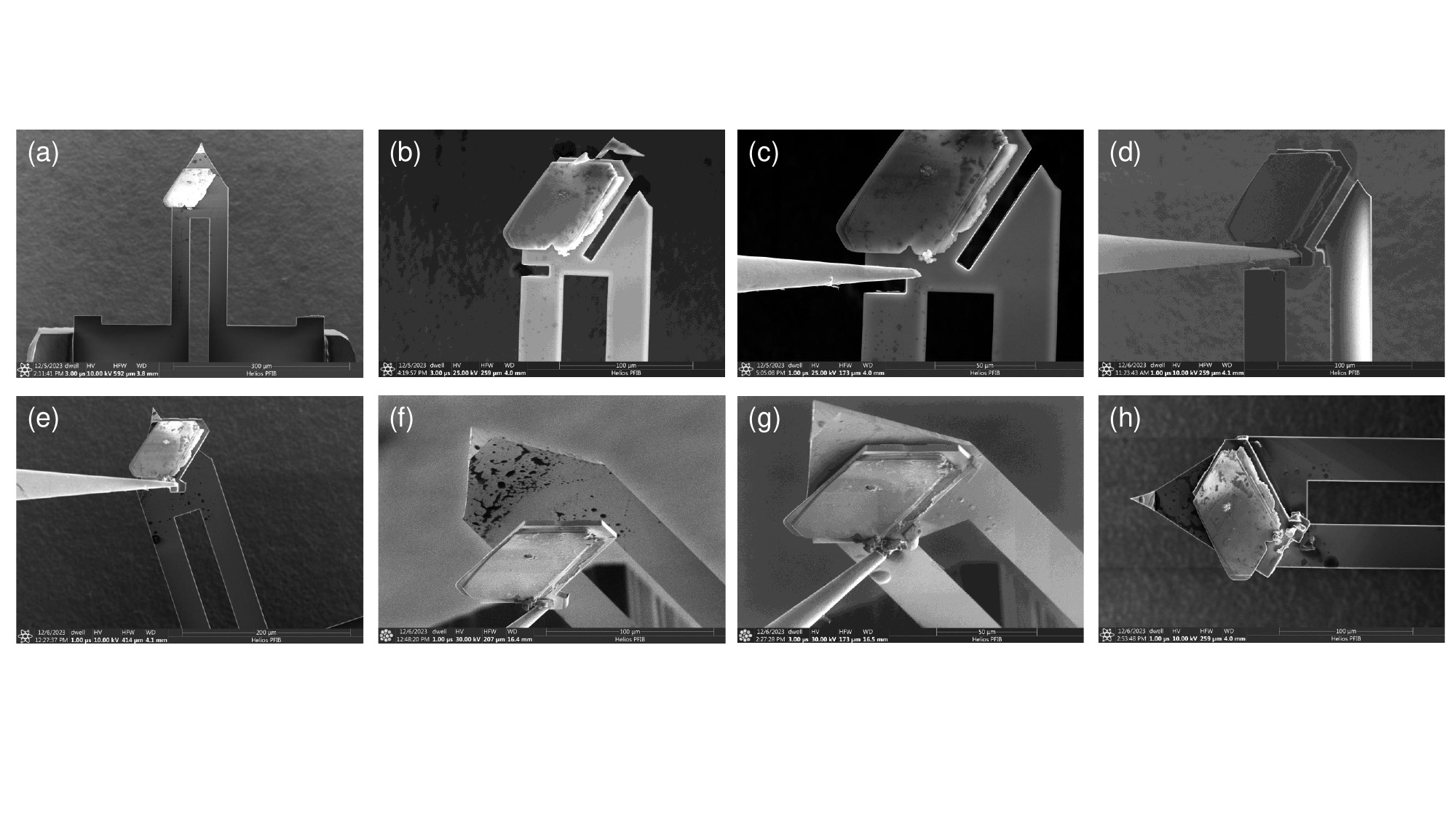}
\caption{\textbf{SEM images of the remounting procedure for $\phi$-dependent magnetotropic measurements.} 
(a) The \rucl crystal mounted on the tip of an Akiyama probe \cite{Akiyama2010}. (b) After permanent adhesion of the crystal to the lever, focused ion beam (FIB) milling of the cantilever around the sample. (c) Tungsten needle welded to the silicon underneath the crystal. (d) FIB milling to isolate the silicon below the crystal from the rest of the Akiyama probe. (e) The needle carrying the sample approaches a new cantilever. (f) The sample positioned $\sim$100~$\mu$m above the new cantilever. (g) The silicon below the crystal is welded with platinum onto the new cantilever. (h) The FIB is used to detach the needle.
}
\label{fig:S2}
\end{figure*}

The \rucl crystal was placed on the cantilever and fixed in place using a small drop of Apiezon N grease. Once the grease is exposed to the electron beam inside the scanning electron microscope (SEM) chamber, it hardens and permanently fixes the sample in place on the cantilever. In order to rotate the crystal about the axis normal to the surface of the cantilever, the FIB was used to cut through the original cantilever that the crystal was permanently fixed to, and transfer it to another cantilever for measurement (\autoref{fig:S2}). For all future measurements, the silicon below the crystal was never permanently fixed to the cantilever; as the grease softens at room temperature, the crystal and silicon can easily be rotated by pushing on the side of the silicon with an eyelash. 
This allowed for repeated reorientation of the sample for each $\phi$ measurement, without touching the crystal and thus, minimizing the risk of inducing stacking disorder.

\section{Magnetotropic susceptibility measurements}

\subsection{Across the AFM phase boundary}
\label{sec:AFM}

The jump in the magnetotropic susceptibility at a phase transition was derived in \citet{Modic2018}. The change in $k$ across a phase boundary depends both on the change in the specific heat $\Delta C$ and the slope of the AFM phase boundary squared.
\begin{equation}
\Delta k = -\frac{\Delta C}{T_N}\left(\frac{\partial T_N}{\partial\theta}\right)^2_B.
\label{eq:deltak}
\end{equation}
In RuCl$_3$, this leads to a strongly-enhanced thermodynamic signature of the phase transition at high fields applied near the $c^\star$-axis \cite{Modic2020}. The technique is therefore exceptionally suited to highlight magnetic phase transitions associated with $R\bar{3}$ domains if they exist, and more generally, to detect phase transitions in anisotropic quantum magnets.

For a minority structural fraction $f$ within an otherwise majority-phase crystal, only that fraction of the sample contributes to the entropy change at its own transition, so the observed jump becomes
\begin{equation}
\Delta k_\mathrm{obs} = -\frac{f\,\Delta C}{\tilde{T}_N}\left(\frac{\partial \tilde{T}_N}{\partial\theta}\right)^2_B,
\label{eq:deltakminor}
\end{equation}
where $\tilde{T}_N$ denotes the transition temperature of the minority phase. Because $\Delta k_\mathrm{obs}$ scales with the square of the angular slope of the phase boundary, a small phase fraction $f \ll 1$ can nonetheless produce an observable, even prominent, signature wherever $\left(\partial \tilde{T}_N/\partial\theta\right)_B$ is large---compensating for the smallness of $f$.

This is the situation realized in RuCl$_3$. For in-plane fields, the AFM phase boundary is close to an angular extremum, so $\left(\partial T_N/\partial\theta\right)_B \to 0$ and both the majority- and minority-phase transitions produce only weak features, consistent with the minority $R\bar{3}$ transitions being only marginally resolvable in the field-derivative of $k$ for purely in-plane sweeps (\autoref{fig:derivative}). As the field is tilted toward $c^\star$, $\left(\partial T_N/\partial\theta\right)_B$ increases sharply for both phases, and the minority-phase signal becomes comparably prominent to the majority-phase signal despite $f \ll 1$.

This enhancement mechanism, being a property of Eq.~\ref{eq:deltakminor} alone, applies identically regardless of whether the transition in question belongs to the majority or minority structural phase; it is not specific to the $R\bar{3}$ fraction, nor does it depend on any detail of the minority phase's magnetic structure beyond the steepness of its own phase boundary.

\subsection{Selection of the critical fields and reproducibility}

\begin{figure}[ht]
  \centering
  \includegraphics[width=0.75\columnwidth, trim=0cm 0cm 0cm 0cm, clip=true]{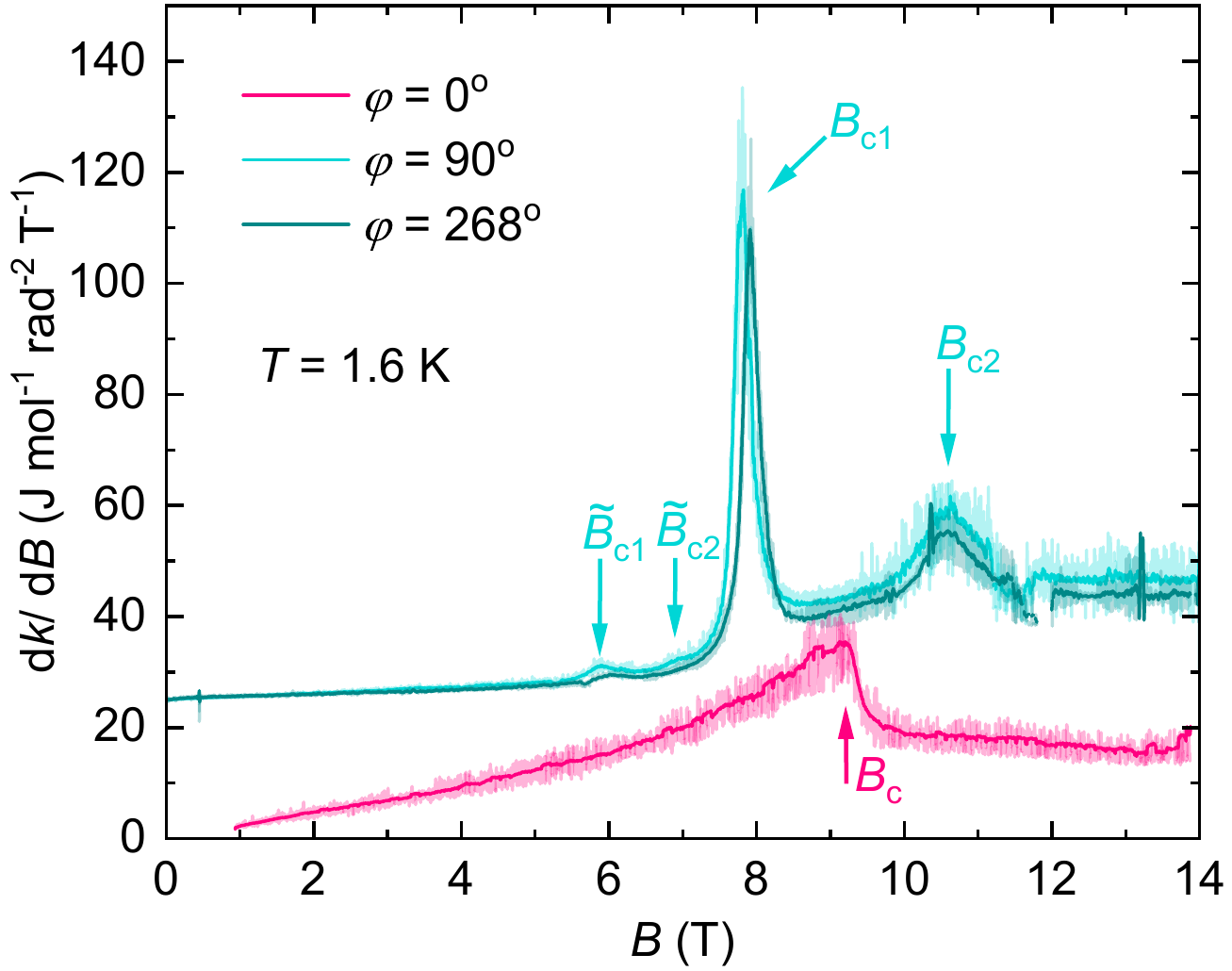}
  \caption{
  \textbf{First field-derivative of the magnetotropic susceptibility, $\mathrm{d}k/\mathrm{d}B$, with the magnetic field applied along the $a$, $b$, and $-a$ directions.} For each angle ($\phi = 0^{\circ}, 90^{\circ}, 268^{\circ}$), faint-colored curves represent the raw data and dark-colored curves represent the smoothed data. The first-order transition $B_\text{c1}$ and the suppression of antiferromagnetism $B_\text{c2}$ of the \mono structure are clearly resolved. The antiferromagnetic transitions of the rhombohedral structure, $\tilde{B}\text{c1}$ and $\tilde{B}\text{c2}$, as previously reported by \citet{Tanaka2022} and others \cite{Balz2021,Balz2019,bachus2021angle,Bachus2020}, are indicated. For $B \parallel b$ ($\phi = 0^\circ$), a single transition is observed at $B_\text{c} \simeq 9.2$~T.
  }
  \label{fig:derivative}
  \end{figure}

\autoref{fig:derivative} presents the field-derivative of the magnetotropic susceptibility, $\mathrm{d}k/\mathrm{d}B$, as a function of the applied in-plane magnetic field $B$. The field was applied along two symmetric in-plane directions: the $a$ and $-a$ axes of the crystal at $\phi = 90^{\circ}$ and $\phi = 268^{\circ}$, respectively. These two curves overlap nicely across the entire field range, suggesting that strain induced by thermal cycling is negligible.  In addition to the peaks in $\mathrm{d}k/\mathrm{d}B$ associated with the \mono structure at 7.9~T and 10.7~T, a faint feature is also visible at 5.9~T and even fainter at 6.9~T. This suggests that a small portion of the crystal has \rhom symmetry. As explained above in \autoref{sec:AFM}, only a small change in the slope of $k$ versus field is observed at the AFM transition under in-plane field configurations. For this field configuration, we rely on the field-derivative data shown in \autoref{fig:derivative} to identify the positions of the critical fields taken as the maximum in the derivative.

\begin{figure}[H]
  \centering
  \includegraphics[width=0.75\columnwidth, trim=5cm 0cm 5cm 0cm, clip=true ]{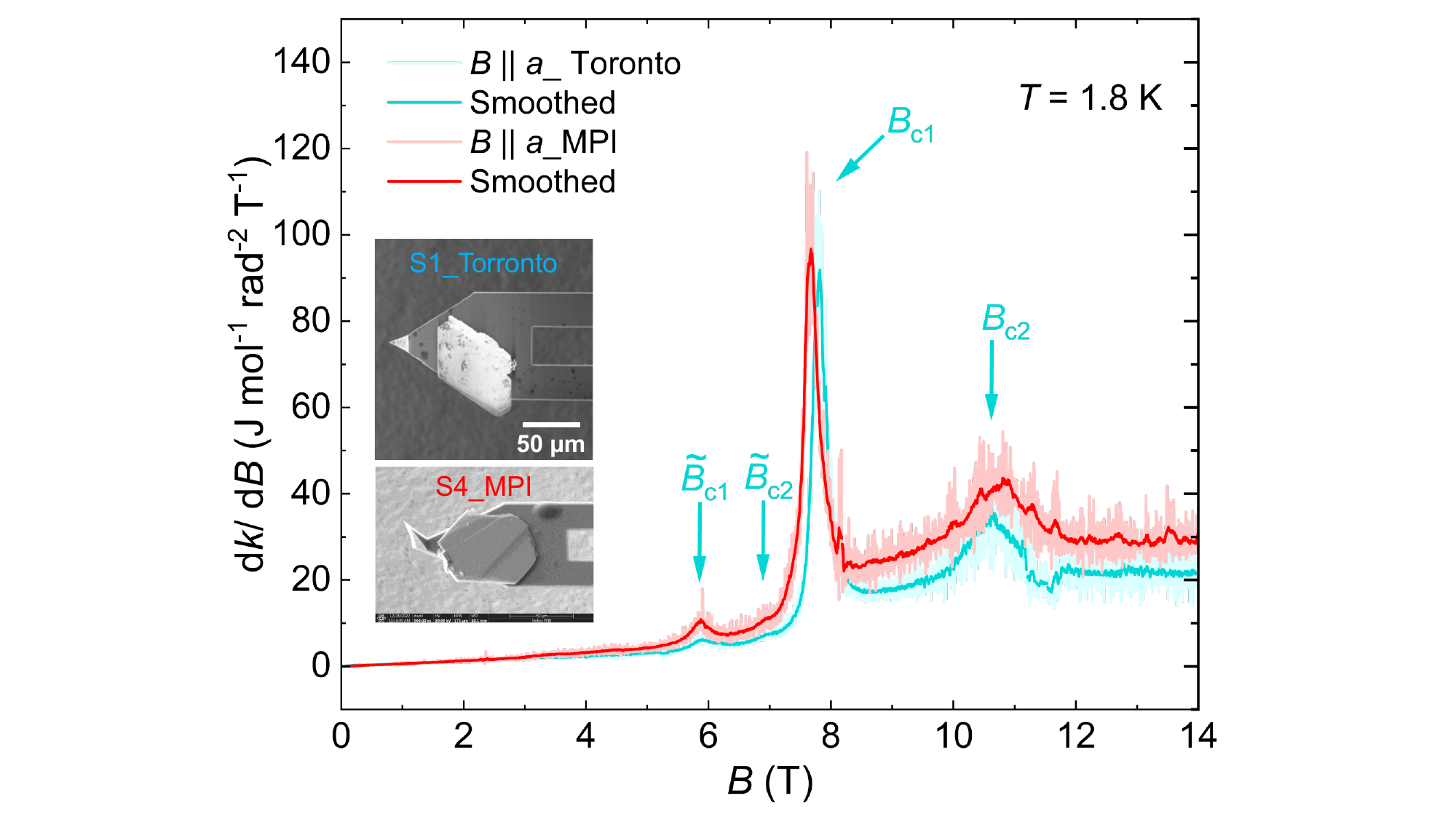}
  \caption{\textbf{First derivative of the magnetotropic susceptibility, $\mathrm{d}k/\mathrm{d}B$, with the magnetic field applied along the $a$ direction for two samples from different sources.} Sample S1 is used in the main study, and Sample S4 is from the batch of samples used in \cite{Modic2020, Modic2018}. The inset shows the crystals mounted on the cantilever.
}
  \label{fig:Bsweep_s1s4}
  \end{figure}

\autoref{fig:Bsweep_s1s4} shows the field derivative of the magnetotropic susceptibility, $\mathrm{d}k/\mathrm{d}B$, for Sample S1, which is used in the main text, and for Sample S4, which is obtained from an independent batch supplied by the Max Planck Institute---also used in Refs.~\cite{Modic2018,Modic2020}. The sample masses are within a factor of 2 of each other. They were both measured at $T\sim1.8$~K with $B\parallel a$. In both datasets, all four critical fields are observed in close agreement, especially considering the highly-anisotropic nature of \rucl; small field-angle misalignment may account for the small differences observed.  The overall change in $k$ at each transition is roughly the same across both datasets, consistent with the interpretation that small samples are predominantly \mono, displaying larger signatures at $B_\text{c1}$ and $B_\text{c2}$. The weaker  signatures observed at $\tilde{B}_\text{c1}
$ and $\tilde{B}_\text{c2}
$ are consistent with a small \rhom fraction in both samples. That the four critical fields appear at the same values across two datasets demonstrates that the magnetic phase boundaries ascribed to the \rhom and \mono fractions are highly reproducible across samples from different growth batches in different groups.

\begin{figure}[ht]
 \centering
 \includegraphics[width=1\columnwidth, trim=0cm 0cm 0cm 0cm, clip=true]{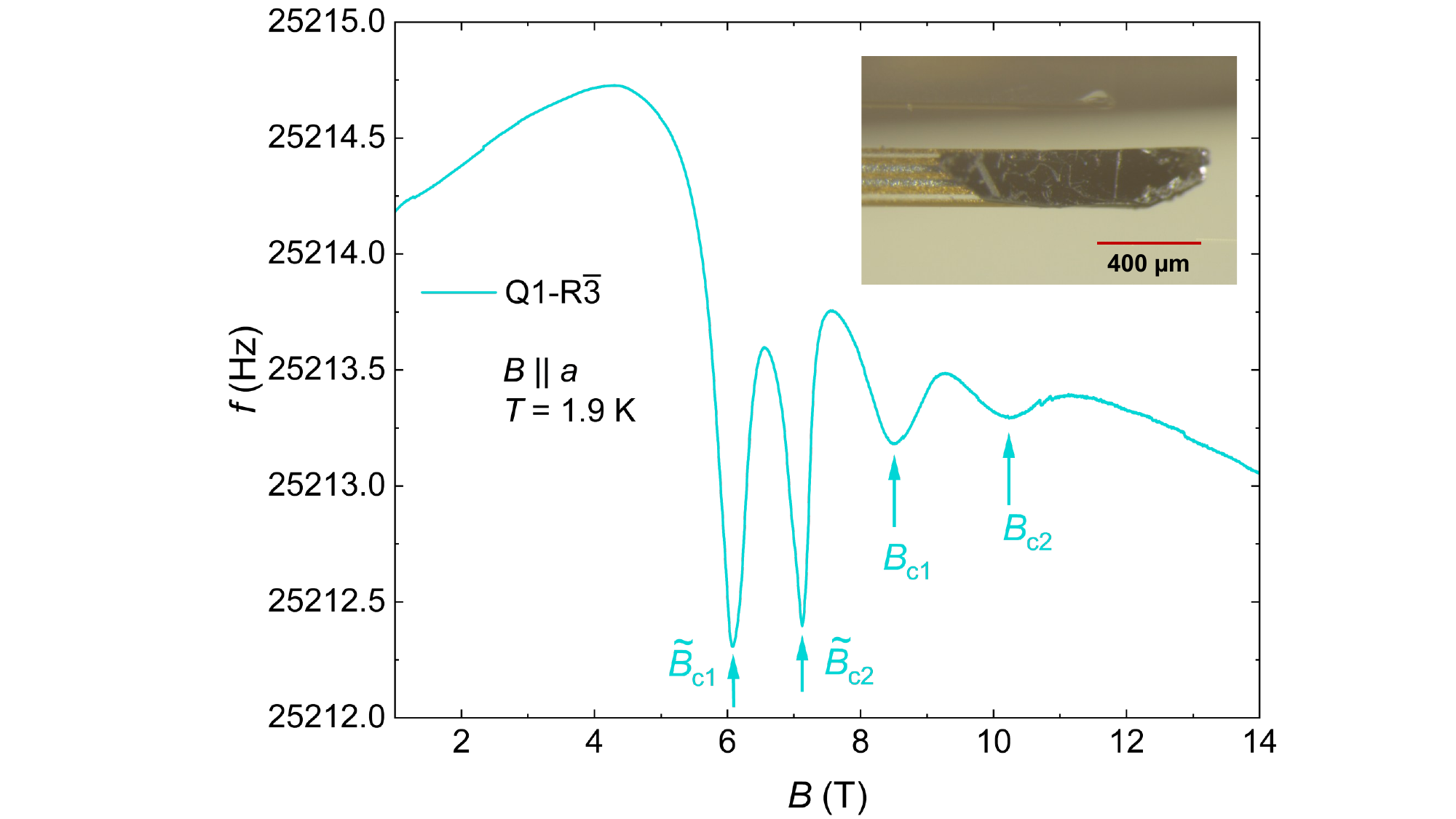}
  \caption{\textbf{Magnetotropic susceptibility on a large sample ($\sim40$ $\mu$g).} Measured frequency shift (directly proportional to the magnetotropic susceptibility) versus magnetic field aligned along the $a$-axis at $T = 1.9$ K. Data were taken on Sample Q1-\rhom, which is $\sim$1000 times larger in mass than Sample S1 and S4 measured in this study. The inset shows the crystal mounted on the qPlus sensor---a single-pronged tuning fork. Four transitions are identified as dips in the measured resonant frequency: $\tilde{B}_\text{c1}$ corresponds to entry into the $ZZ2$ phase of the \rhom structure. $\tilde{B}_\text{c2}$ corresponds to suppression of AFM order associated with the \rhom structure. $B_\text{c1}$ and $B_\text{c2}$ correspond to the analogous transitions associated with the $X$ phase of the \mono structure.
}
 \label{fig:Bsweep_q1}
\end{figure}

To further investigate the assignment of the weaker, lower-field features to a minority \rhom fraction, we performed complementary measurements on a \rhom-dominant crystal (Sample Q1-\rhom) using a \href{https://qplussensor.com/}{qPlus sensor}. Sample Q1-\rhom is $\sim$1000 times larger in mass than Samples S1 and S4 (values are listed in \autoref{tableS1}). \autoref{fig:Bsweep_q1} shows the resonance frequency as a function of field measured at $T\sim1.9$~K with $B\parallel a$. The two largest anomalies appear at the fields marked by $\tilde{B}_{\text{c1}}$ and $\tilde{B}_{\text{c2}}$, followed by weak features at higher fields, $B_\text{c1}$ and $B_{\text{c2}}$. The higher field transitions correspond to the values identified in this work as the main magnetic transitions of the \mono phase, while the lower field transitions coincide with the known magnetic phases of the \rhom phase. These four distinct anomalies, with the amplitude of the transitions swapped compared to the smaller crystals measured in \autoref{fig:Bsweep_s1s4}, demonstrate that: i) larger samples are dominated by the \rhom phase and ii) the two transitions occurring at lower fields $\tilde{B}_{\text{c1}}$ and $\tilde{B}_{\text{c2}}$ result from the magnetic phases associated with that \rhom fraction. The higher-field transitions $B_\text{c1}$ and $B_{\text{c2}}$ are clearly visible in the raw data,  showing that the \mono fraction is still significant, but greatly reduced compared with the \rhom fraction. 

\subsection{In-plane anisotropy of the N\'{e}el temperature at fixed fields}
\begin{figure*}[ht]
  \centering
  \includegraphics[width=1\columnwidth]{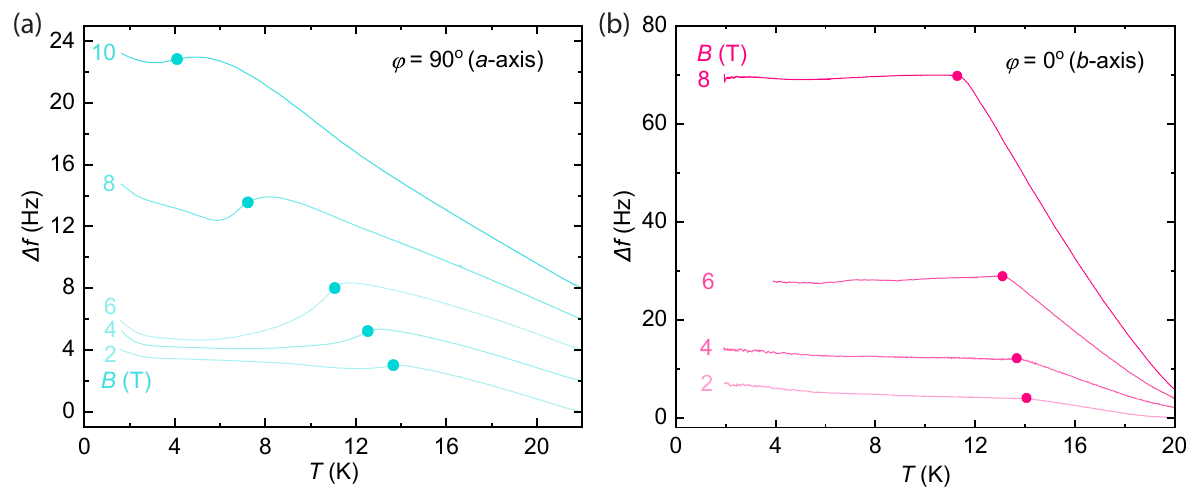}
  \caption{
  \textbf{Temperature dependence of the frequency shift (directly proportional to $k$) at fixed magnetic fields.}
  (a) Temperature sweeps from 22 to 1.5~K measured at discrete fields from 2–10~T along the $a$-axis. (b) Temperature sweeps from 20 to 1.5~K in fixed magnetic fields from 2–8~T along the $b$-axis. Colored circles indicate the transition temperature into the AFM phase.
}
  \label{fig:S4}
  \end{figure*}

\autoref{fig:S4} presents the temperature dependence of the magnetotropic response measured at different magnetic field strengths, from which the antiferromagnetic phase boundary shown in Figure 3 of the main text is determined. For fields applied along the $a$-axis (panel a), the magnetic field required to suppress AFM order is larger than for $B||b$ (panel b). Consequently, $T_\text{N}$ remains observable up to 10~T for $B||a$. For $B||b$, $T_\text{N}$ has a weaker field dependence below 8 T compared to $B||a$, however, by 10 T, AFM is already suppressed for $B||b$. This demonstrates consistency with the critical fields observed in the field sweeps shown in Figure 3b of the main text.

\subsection{Field angle evolution of the AFM phases}

In fully mapping out the AFM phase diagram of a \mono-dominant \rucl crystal at low temperatures, we uncovered a distinct magnetic phase for $B||a$. While magnetotropic measurements cannot distinguish the magnetic texture of the $X$ phase, we find similarities between it and the $ZZ2$ phase found in \rhom-dominant samples. In both cases, the intermediate phase exists for $B||a$ and is limited to a narrow field-angle range within $\sim$30 degrees of the $ac^\star$-plane. Note that in some previous studies, the $ZZ2$ phase has been reported to exist for all in-plane field angles \cite{Tanaka2022, bachus2021angle}. However, we believe that this arises from rotated twin domains, largely in samples containing mixtures of the \rhom and \mono phases. 

\begin{figure*}[ht]
  \centering
  \includegraphics[width=0.75\columnwidth]{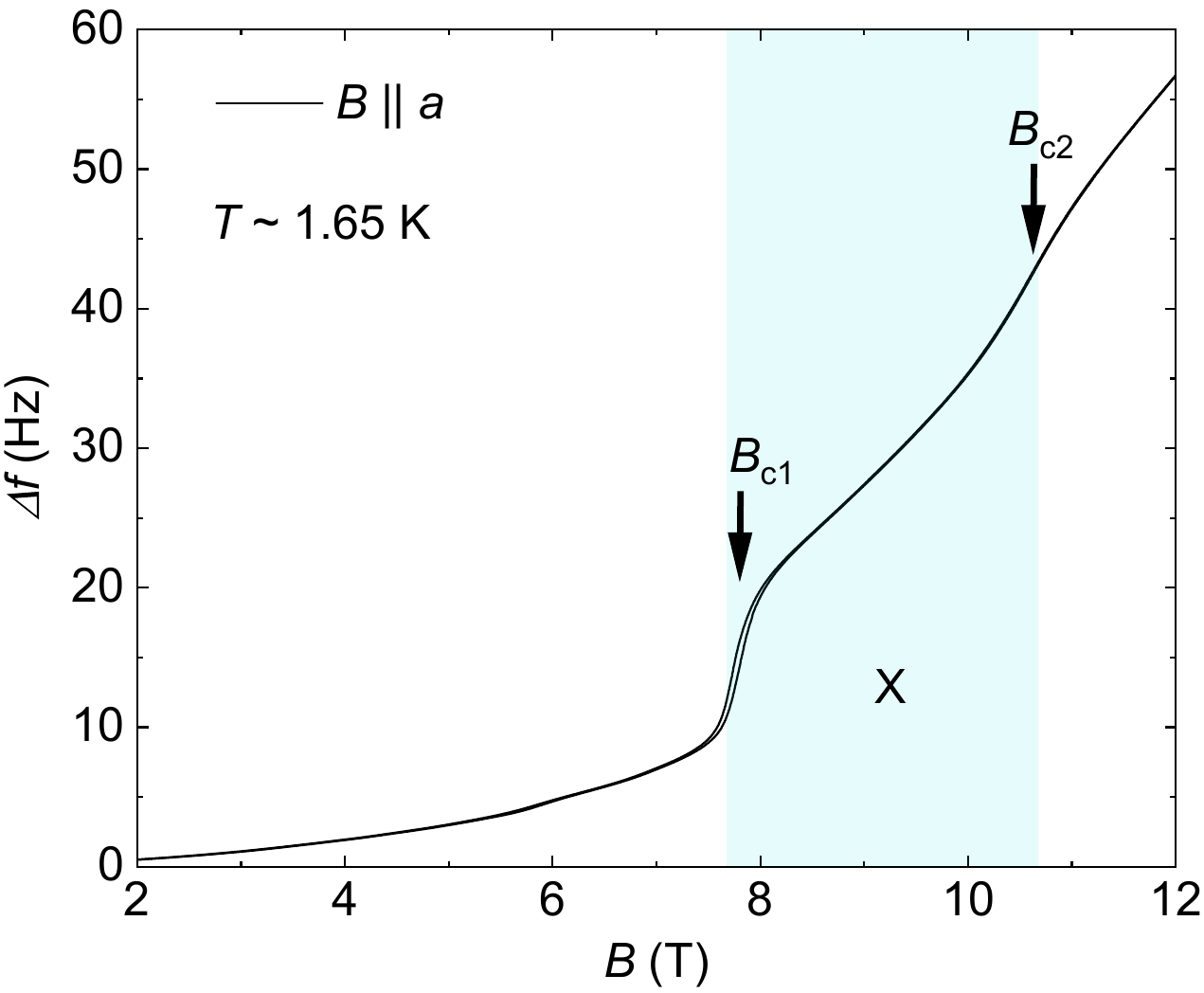}
  \caption{\textbf{Hysteresis at the first-order phase transition into the $X$ phase.} Magnetic-field dependence of the frequency shift \(\Delta f\) (proportional to $k$) at \(T\approx1.65\) K for magnetic field increased and decreased along the $a$-axis. The data were taken on Sample S1. Two transitions are observed: \(B_{c1}\) is a first-order transition with a narrow hysteretic region marking entry into the $X$ phase, and \(B_{c2}\) is a second-order transition marking the upper bound of the intermediate $X$ phase (cyan-shaded region).
  }
  \label{fig:Bsweep hysterises}
\end{figure*}

As a further check on the thermodynamic character of the $X$ phase, we performed increasing and decreasing field sweeps through the transition and observed a small hysteretic region surrounding $B_\text{c1}$ (\autoref{fig:Bsweep hysterises}). This hysteresis confirms that $B_\text{c1}$ is a genuine first-order phase transition: it separates two distinct ordered states, ruling out a broadened crossover that could arise from a distribution of local transition fields. Because $B_\text{c1}$ is the dominant feature resolved---orders of magnitude stronger than the minority $R\bar{3}$ signatures identified elsewhere in this work (\autoref{fig:derivative})---the first-order character is intrinsic to the bulk $C2/m$ phase.

\begin{figure*}[ht]
  \centering
  \includegraphics[width=1\textwidth, clip=true ]{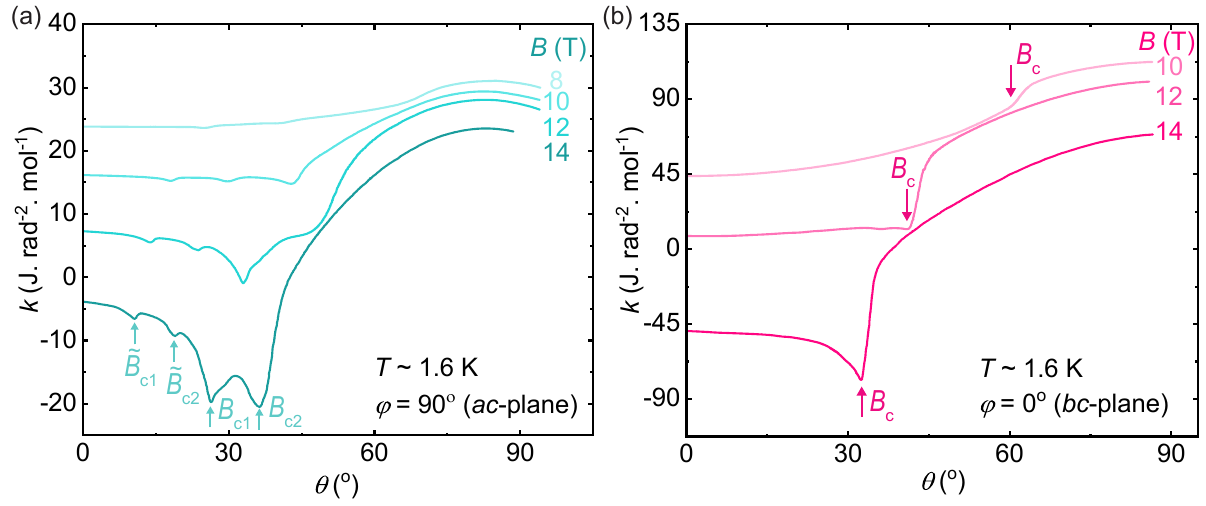}
  \caption{
  \textbf{Angular dependence of the critical fields at low temperature.} 
  (a) Angular dependence of the magnetotropic response in the $ac$-plane at $T = 1.6$ K. Arrows mark the critical field orientation where the system crosses the AFM phase boundary. $\theta=0^\circ$ is defined as the field aligned with the $c^\star$-axis. For each of the curves shown at a fixed field value, AFM order is observed at $\theta=0^\circ$ and for nearby angles up to the marked transitions $B_\text{c}$, beyond which a correlated Kitaev paramagnetic state is reached.
  (b) Angular dependence in the $bc$-plane at $T = 1.6$ K measured at different fixed magnetic fields. Arrows indicate the critical angle at which AFM order is fully suppressed as the field is rotated away from the $c^\star$-axis.
}
  \label{fig:S5}
  \end{figure*}

\autoref{fig:S5} presents the angular dependence of the magnetotropic response measured at $T \approx 1.6$~K for magnetic fields ranging from 8 to 14~T rotated in the $ac$- and $bc$-planes. As the field is rotated from the in-plane direction ($\theta = 90^\circ$) towards the out-of-plane direction ($\theta = 0^\circ$), the critical field required to suppress the ordered state increases significantly \cite{Modic2020,Zhou2023,Gordon2019a}. Consequently, the system remains in the ordered phase over a narrower angle range as magnetic field increases. Field-angle scans provide an alternative route to pierce through the AFM phase at the available field strengths (up to 14 T) where the boundary of the AFM phase changes sharply with field-orientation (\textit{i.e.} at fixed fields closer to the $c$-axis where $T_\text{N}$ exhibits large changes with field angle). This leads to enhanced signatures in $\Delta k$ at the phase transitions, highlighting the four critical fields. More specifically, the two transitions associated with the \rhom fraction, $\tilde{B}_{c1}$ and $\tilde{B}_{c2}$, which were hardly visible under in-plane magnetic fields, are strongly enhanced under out-of-plane magnetic fields. This is a key advantage of resonant torsion that is central to this study---the ability to track the phase transitions associated with the small \rhom fraction (towards the known in-plane field values) relies on the sensitivity of resonant torsion to thermodynamic signatures of anisotropic ordered phases.    

In \autoref{fig:S5}(a), field-angle dependent scans of $k$ at 14~T reveal four distinct minima, highlighting the presence of both the $ZZ2$ phase and the $X$ phase in the $ac^\star$-plane. Arrows in both panels indicate the critical fields associated with the \mono phase ($B_{c1}$ and $B_{c2}$), as well as those attributed to the \rhom contribution ($\tilde{B}_{c1}$ and $\tilde{B}_{c2}$). In panel b, for measurements in the $bc^\star$-plane, at a fixed field as $\theta$ increases, a single transition corresponding to the suppression of AFM associated with the \mono phase is observed at $B_\text{c}$. The small shoulder appearing at lower angles next to $B_\text{c}$ for $B = 12, 14$ T is likely the critical field associated with the small \rhom fraction. 

\begin{figure*}[ht]  
  \centering
  \includegraphics[width=1\textwidth, trim=0cm 0cm 0cm 0cm, clip=true]{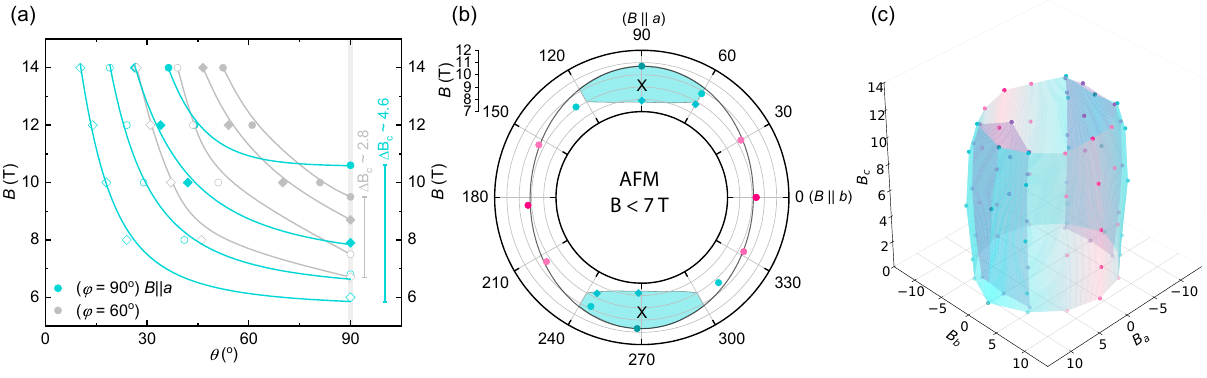}
  \caption{
  (a) Critical fields extracted from angle-sweep measurements at $\phi = 60^\circ$ and $90^\circ$ for different applied magnetic field strengths. The critical field window of the transitions at $\phi = 60^\circ$ is approximately half that observed for $\phi = 90^\circ$.
  (b) Polar representation of the angular dependence of the critical fields ($B_c$), highlighting the pronounced twofold in-plane anisotropy of the AFM phase boundary. The field-induced $X$ phase is confined to a narrow angular range in the vicinity of the $a$-axis direction.
  (c) Three-dimensional representation of the AFM phase boundary of monoclinic \rucl, including the intermediate $X$ phase.
}
  \label{fig:all critical fields}
\end{figure*}

The angle-dependent critical fields are summarized in \autoref{fig:all critical fields}. The first panel compares the field-angle dependence of the transitions for both the \rhom and \mono structures to illustrate how the $ZZ2$ and the $X$ evolve as the magnetic field is rotated in two planes that are $\sim30^\circ$ from one another (cyan versus gray curves). The field-angle region over which these phases exist significantly narrows as magnetic field is applied further from the $ac^\star$-plane. As $\phi$ changes from $90^\circ$ ($B||a$) to $\phi = 60^\circ$, the field window over which the field-induced phase transitions observed for in-plane fields (at $\theta = 90^\circ$) reduces by a factor of roughly 2.  This behavior is consistent with previous studies showing that the intermediate zigzag phase rapidly diminishes upon rotating an in-plane magnetic field away from the $a$-axis \cite{Bachus2020,bachus2021angle,Lefrancois2023}.

The critical fields intrinsic to the \mono structure for in-plane fields are shown as a polar plot in \autoref{fig:all critical fields}b. 
This captures the pronounced twofold anisotropy expected for the \mono structure. It also highlights the systematically higher critical fields compared to the \rhom phase \cite{Tanaka2022,Balz2019,Balz2021,Cen2025,Sears2017,Namba2024}. 
In panel c, the full $\theta$- and $\phi$-dependence of the AFM phases associated with the \mono structure are shown at $T = 1.5$ K and up to 14 T. The intermediate $X$ phase is confined to fields applied within $\pm 30^\circ$ from the $ac$-plane.



\setcounter{NAT@ctr}{0}
\clearpage
\putbib[RuCl3_refs,RuCl3_refs_2]

\end{bibunit}
\end{document}